\documentclass[a4paper, 10pt]{article}
\usepackage[utf8]{inputenc}
\usepackage{lmodern}
\usepackage{amsmath}
\usepackage{fixmath}
\usepackage{exscale}
\usepackage{accents}
\usepackage{bigints}
\usepackage{physics}
\usepackage{commath}
\usepackage{float}
\usepackage{subfigure}
\usepackage{caption}
\usepackage{physics}
\usepackage{graphicx}
\usepackage{upgreek}
\usepackage{textcomp}
\usepackage{amssymb}
\usepackage{bbold}
\usepackage{stackengine}
\usepackage{tabularx}
\usepackage[title]{appendix}
\usepackage{stmaryrd}
 \usepackage[top = 0.5in, left = 0.75 in, right = 0.75 in, bottom = 0.8 in]{geometry}
 \allowdisplaybreaks

\usepackage[authoryear,round]{natbib}
\usepackage{bm}
\usepackage{authblk}

\usepackage{moreverb}

\numberwithin{equation}{section}

\usepackage{bbm}

\newcommand{\mbf}[1]{\mathbf{#1}}
\newcommand{\bsym}[1]{\boldsymbol{#1}}
\newcommand{\bbm}[1]{\mathbbm{#1}}
\newcommand{\mcal}[1]{\mathcal{#1}}
\newcommand{\bbbm}[1]{\bsym{\bbm{#1}}}

\newcommand{\jump}[1]{\llbracket {#1} \rrbracket}

\usepackage{setspace}
\usepackage[colorlinks=true, urlcolor=blue, pdfborder={0 0 0},citecolor=blue]{hyperref}

\newcommand\BibTeX{{\Today's agenda is to discuss the recent modifications made to the draft, specifically regarding the curl-free magnetic field's condition and its integration into the current manuscript.rmfamily B\kern-.05em \textsc{i\kern-.025em b}\kern-.08em
T\kern-.1667em\lower.7ex\hbox{E}\kern-.125emX}}

\date{}
\title{
A fully--coupled nonlinear magnetoelastic thin shell formulation}
\author[1]{Abhishek Ghosh}

\author[1]{Andrew McBride}
\author[2]{Zhaowei Liu}
\author[3]{Luca Heltai}
\author[1,4]{Paul Steinmann}
\author[1]{Prashant Saxena\thanks{Corresponding author: \it{prashant.saxena@glasgow.ac.uk}}}
\affil[1]{\small Glasgow Computational Engineering Centre, James Watt School of Engineering, University of Glasgow, Glasgow, G12 8LT, United Kingdom}
\affil[2]{\small College of Mechanics and Materials, Hohai University, Nanjing, 211100, China}
\affil[3]{\small SISSA (International School for Advanced Studies), Via Bonomea 265, 34136 Trieste, Italy}
\affil[4]{\small Institute of Applied Mechanics, Friedrich-Alexander Universit\"at Erlangen-N\"urnberg, D-91052, Erlangen, Germany}

\begin{document}
\maketitle
\begin{abstract}
A geometrically exact dimensionally reduced order model for the nonlinear deformation of thin magnetoelastic shells is presented.
The Kirchhoff--Love assumptions for the mechanical fields are generalised to the magnetic variables to derive a consistent two-dimensional theory based on a rigorous variational approach.
The general deformation map, as opposed to the mid-surface deformation, is considered as the primary variable resulting in a more accurate description of  the nonlinear deformation.
The commonly used plane stress assumption is discarded due to the Maxwell stress in the surrounding free-space requiring careful treatment on the upper and lower shell surfaces.
The complexity arising from the boundary terms when deriving the Euler--Lagrange governing equations 
is addressed via a unique application of Green's theorem.
The governing equations are solved analytically for the problem of an infinite cylindrical magnetoelastic shell. This clearly demonstrates the model's capabilities and provides a physical interpretation of the new variables in the modified variational approach.
This novel formulation for magnetoelastic shells 
serves as a valuable tool for the accurate design of thin magneto-mechanically coupled devices.

\end{abstract}

\textbf{Keywords:} Nonlinear magnetoelasticity; Magnetoelastic shells; Kirchhoff--Love; Large deformation


\section{Introduction}
Large deformation of thin structures made from soft rubbers or elastomers is critical for numerous engineering components, including tyres, airbags, air springs, buffers, pneumatic actuators, and soft grippers \citep{galley2019pneumatic, hao2017modeling}.
The analysis of slender structures undergoing large deformation, such as rods, membranes, plates, and shells, in which one or more characteristic dimensions are negligible compared to the others, is challenging. 
They exhibit both material and geometric nonlinearities often leading to instabilities. Slender structures are generally modelled as lower-dimensional manifolds embedded in three-dimensional space with appropriate kinematic simplifications \citep{Niordson1985, Simo1989OnSR}.
Novel developments in smart materials with multi-physics coupling have led to a dramatic increase in technological applications in soft robotics, actuators, and sensors.
These materials often rely on a non-mechanical stimulus from electric, magnetic, thermal or chemical fields \citep{Mark_R_Jolly_1996, McKay_2010, kim-swell2012,  Sheng2012} and are difficult to model due to their complex physics.
Of particular relevance is magneto-mechanical coupling in thin structures due to
the ability to produce
extremely large reversible deformations in a  short time-scale.
The presence of strong magneto-mechanical coupling in some manufactured materials, such as magnetorheological elastomers (MREs) \citep{Mark_R_Jolly_1996}, has the potential to underpin future engineering and technological applications, for example, in micro-robotics \citep{Hu2018,Ziyu2019}, as sensors and actuators \citep{Bose2012,Psarra2017}, in active vibration control \citep{Ginder2000}, and as waveguides \citep{Saxena2018,Karami2019}.

Magnetoelastostatics concerns the analysis of suitable phenomenological models to describe the equilibrium of deformable solids associated with multifunctional processes involving magnetic and elastic effects. The main constituent of the theory is the coupling between elastic deformation and magnetisation in the presence of externally applied mechanical and magnetic force fields. The magnetoelastic coupling  occurs in response to a phenomenon involving reconfigurations of small magnetic domains.  This is observable as a continuum vector field emerging from an averaging of microscopic and distributed subfields. Thus, the imposition of a magnetic field also induces a deformation of the material in addition to the magnetic effects caused by the traditional mechanical forces.
With a rich history spanning six decades \citep{ Tiersten1964, Brown1966, Maugin1972,Maugin1988, DeSimone2002, Kankanala2004, Ogden2004, KEIP2019805, Sharma2020, MORENOMATEOS2023104742}, the mathematical and computational modelling of magnetoelasticity continues to be an active area of research. 
The coupling between magnetic fields and mechanical deformations of a shell structure introduces additional complexities compared to purely mechanical or electromagnetic analyses, making the modelling and solution process considerably more challenging. 

Magneto-active soft materials are broadly divided into two sub-classes based on the type of embedded particles: soft-magnetic soft materials (SMSMs) and hard-magnetic soft materials (HMSMs). 
SMSMs contain particles with low coercivity, such as iron or iron oxides, and their magnetization vector varies under external magnetic loading. 
They are often modelled as three-dimensional solid continua \citep{DANAS2012120, Saxena2013, Ethiraj2016, MEHNERT2017117, MUKHERJEE2020103380, BUSTAMANTE2021103429, Akbari_2021, HU2022111310}.
HMSMs consist of particles with high coercivity, such as CoFe$_2$O$_4$ or NdFeB. The magnetisation vector, or remnant magnetic flux, of HMSMs remains unchanged over a wide range of applied external magnetic flux \citep{LEE2020108921, Schumann2021, MORENOMATEOS2023105232}.
The viscoelastic material behaviour of HMSMs significantly affects the magnetic actuation behaviour of hard-magnetic soft actuators \citep{LUCARINI2022111981, Sharma2023, Anand2023}.

Motivated by the  need  to model thin magnetoelastic structures, \cite{steigmann2004} presented a dimensionally reduced-order model for thin magneteolastic membranes.
\cite{Barham_2008} investigated the limit point instability for a finitely deforming circular magnetoelastic membrane in an axisymmetric dipole field under  one-way magneto-mechanical coupling.
This analysis was extended by \cite{REDDY2017248, REDDY2018203, SAXENA2019250, Ali2021Magmembrane, MISHRA2023104368} to study wrinkling, bifurcation, and limit point instabilities in axisymmetric inflating magnetoelastic membranes. 
However, a shell theory for fully-coupled  magnetoelasticity that can account for bending resistance is still lacking.
An overview of the classical shell theory is given, for example, in \citet{Simo1989OnSR, Cirak2000, kiendl2009isogeometric} or in the books by \citet{Basar1985, Niordson1985, Blaauwendraad2014}. 
When modelling physical phenomena on curved surfaces, defining geometric quantities (normal vectors, curvatures, etc.) and differential surface operators (gradients, divergence, etc.) is crucial 
\citep{Paul_2015}.

Reduced-order theories for hard-magnetic linear and nonlinear beams \citep{WANG2020104045, YAN2022111319}, and rods \citep{SANO2022104739} have been derived based on the three-dimensional model presented in \cite{ZHAO2019}. 
These studies involved a dimensional reduction procedure on the three-dimensional magneto-elastic energy, assuming  reduced kinematics based on the Kirchhoff–Love assumptions \citep{Niordson1985}. 
 \cite{Green1983} focused on the nonlinear and linear thermomechanical theories of deformable shell-like bodies, considering electromagnetic effects. The development was carried out using a direct approach, utilising the two-dimensional theory of directed media known as Cosserat surfaces.
 \cite{Yan2020} studied linear elastic magneto-active axisymmetric shells made of HMSMs.
 They leveraged the coupling between mechanics and magnetism to tune the onset of instability of shells undergoing pressure buckling. 
 Magnetoelastic shell models for axisymmetric deformation and geometrically exact strain measures were compared with experimental results. 
 Their findings demonstrated that the magnetic field can control the critical buckling pressure of highly sensitive spherical shells with imperfections  \citep{ Hutchinson2016, HUTCHINSON2018157}. 
 \citet{PEZZULLA2022104916} performed a dimensional reduction of the three-dimensional magneto-elastic energy contribution presented in \cite{ZHAO2019} by assuming a reduced kinematics according to the Kirchhoff–Love assumptions, and focussing specifically on hard-magnetic, linear-elastic shells.
 Models for non-axisymmetric deformations of magnetoelastic shells have been derived for shallow shells \citep{Seffen_2016, Loukaides2014}. 
 \cite{Dadgar2023} proposed a micropolar-based shell model to predict the deformation of thin HMSMs, incorporating a ten-parameter formulation that considers the micro-rotation of the microstructure with the enhanced assumed strain method to alleviate locking phenomenona. 
 \cite{HoLee2023} have presented a direct two-dimensional formulation to couple non-mechanical stimuli with large deformation of shells. 
 However, despite this wealth of research,  models for general deformation cases in the context of SMSM shells have received limited attention with the form of the coupling between magnetism and mechanics remaining  an open question.

To address the aforementioned shortcoming, a theory is derived to model large deformations of soft-magnetic hyperelastic thin shells using the Kirchhoff-Love assumption for mechanical deformation and a linearly varying magnetic scalar potential across the thickness of the structure. 
The salient features of the theory are the following:
\begin{enumerate}
\item  
In the present work, a derived theory approach is adopted for SMSM shells by considering the total energy of a three-dimensional incompressible magnetoelastic body and its surrounding space as the starting point. 
A two-dimensional system is derived  based on appropriate approximations for a thin shell by incorporating a new set of generalised solution variables in a modified variational setting. 
The magnetic potential in the free-space is treated as an independent solution variable to capture the underlying physics and strongly couple the magnetoelastic interactions between the shell and the free space. 
This approach is required to formulate
appropriate boundary and interface conditions and ensure consistency in the mathematical modelling of the system.
\item In numerical simulations involving thin structures, the common practice is to apply external hydrostatic pressure at the mid-surface. 
In the present derived theory approach, a distinction is made between the
applied pressures at the top and bottom surfaces.
The implications of this departure from the conventional practice are discussed.
\item 
In the present context, obtaining a dimensionally reduced-order theory entails linearly approximating the variation of the magnetic potential across the shell's thickness. This leads to the total potential in the magnetoelastic body adopting a form similar to the Kirchhoff-Love assumption used for the mechanical behaviour of hyperelastic thin shells.
Such an approximation is well-suited for modelling the physics of thin magnetoelastic shells and facilitates mathematical simplifications.
\item The plane-stress assumption, commonly employed in structural mechanics, assumes negligible stresses in the thickness direction of thin plates or shells. 
However, it is not directly applicable to magnetoelastic shells due to the coupling between magnetic field and mechanical deformation. 
This coupling results in three-dimensional stress and strain states, exemplified by the study of an inflating soft magnetoelastic cylindrical shell presented here. 
The plane-stress assumption fails to consider magnetic field-induced stresses in the thickness direction, leading to inaccuracies. 
\item The physical three-dimensional shell is  conceptualised as a stack of surfaces. 
Thereby, the overall deformation of the shell structure is described by the deformed mid-surface position vector, augmented by a term accounting for the through-thickness stretch and the deformed normal. 
Introducing the first variation of the thickness stretch and the deformed normal in the modified variational format
adds richness and complexity to the derivation of the shell system of equations.
Notably, when obtaining a reduced-order model for the thin soft magnetoelastic shell, a unique application of Green's theorem is required. 
The present work addresses this complexity and provides a suitable generalisation by deriving a system of partial differential equations with boundary terms that encompass these effects.
\end{enumerate}
\subsection{Outline}
The paper is organised as follows:  In Section 2, the mathematical preliminaries and the fundamentals of nonlinear magnetoelastostatics 
are introduced. Sections 3.1 and 3.2 define the geometry and kinematics of nonlinear magnetoelastic thin shells, respectively. In Section 3.3, the expressions for the divergence of the total stress tensor and the magnetic induction vector in the shell are provided. These are essential for deriving the equilibrium equations.
Section 3.4 presents the interface condition on the magnetic potential, which is imposed by the continuity of the tangent space components of the magnetic field at the shell boundaries.
Section 4 introduces the variational formulation accounting for
the magnetoelastic body and the corresponding free space under suitable loading situations. Then, in Section 5, a new set of generalised solution variables for a modified variational format suitable for deriving the shell system of equations is introduced.
Sections 5.2, 5.3, and 5.4 demonstrate the contributions of the stress tensors, magnetic induction vector, and external loads to the modified variational setting, respectively.
Section 6 is dedicated to obtaining the governing equations for the system, and in Section 7, an example of an inflating magnetoelastic thin cylinder is illustrated to derive the response equations for a given boundary-value problem using the derived equations.
Finally, in Section 8, concluding remarks  are presented.
\subsection{Notation}
A variable typeset in a normal weight font represents a scalar. A bold weight font denotes a first or second-order tensor. 
A scalar variable with superscript or subscript indices normally represents the components of a vector or second-order tensor. 
Latin indices $i,j,k,\dots$ vary from $1$ to $3$ while Greek indices $\alpha, \beta, \gamma,\dots$, used as surface variable components, vary from $1$ to $2$. Einstein summation convention is used throughout.
$ \mathbf{e}_i$ represent the basis vectors of an orthonormal and orthogonal system in three-dimensional Euclidean space with $x,y$ and, $z$ as its components. 
The three covariant basis vectors for a surface point are denoted as $\mathbf a_i$, where $\mathbf a_\alpha$ are the two tangential vectors and $\mathbf a_3$ as the normal vector with
$\theta^\alpha$ and $\eta$ as the respective coordinate components. 

The comma symbol in a subscript represents the partial derivative with respect to the surface parameters, for example, $A_{,\beta}$ is the partial derivative of $A$ with respect to  $\theta^\beta$. The scalar product of two vectors $\bm{p}$ and $\bm{q}$ is denoted $\bm{p} \cdot \bm{q}$, and the tensor product of these vectors is a second-order tensor $\bm{H}=\bm{p}\otimes\bm{q}$. Operation of a second-order tensor $\bm{H}$  on a vector $\bm{p}$ is given by $\bm{H}\bm{p}$. The scalar product of two tensors, $\bm{H}$ and $\bm{G}$, is denoted $\bm{H} : \bm{G}$. The notation $\norm{\cdot}$ represents the usual (Euclidean) norm 
For a second-order tensor in its component form $\bm{H}=H^{ij}\mathbf a_i\otimes\mathbf a_j$, the component matrix is denoted $\left[H^{ij}\right]$.
Circular brackets $( \,)$ are used to denote the parameters of a function.  
Square brackets $[ \, ]$ are used to group expressions.
If brackets are used to denote an interval, then $(\,)$ stands for an open interval and $[\,]$ is a closed interval. 
\section{Nonlinear Magnetoelastostatics}

A brief review of the key equations of nonlinear magnetoelastostatics is provided, see  \cite{dorfmann2014, Pelteret2020} for further details.
Either the magnetic field, magnetic induction, or the magnetisation vector can be selected as the independent magnetic variable.
The 
present work is established on the variational formulation based on the 
magnetic field.

Consider a magnetoelastic body that occupies the regions ${\mathcal{B}}_{\mathrm{0}} \in \mathbb{R}^3$ and ${\mathcal{B}} \in \mathbb{R}^3$ in its reference and deformed configurations, respectively, with corresponding boundaries denoted as $\partial{\mathcal{B}}_{\mathrm{0}}$ and $\partial\mathcal{B}$.
A point ${\bm{X}}_{\mathrm{B}} \in {\mathcal{B}}_{\mathrm{0}}$ is 
related to a point  $\bm{x}_{\mathrm{B}}\in \mathcal{B}$ through a one-to-one map
${\bm{\chi}}_{\mathrm{B}}( {\bm{X}}_{\mathrm{B}}): {\mathcal{B}}_{\mathrm{0}} \rightarrow \mathcal{B}$.
\begin{figure}[]
  \begin{center}
   \includegraphics[width=0.8\linewidth]{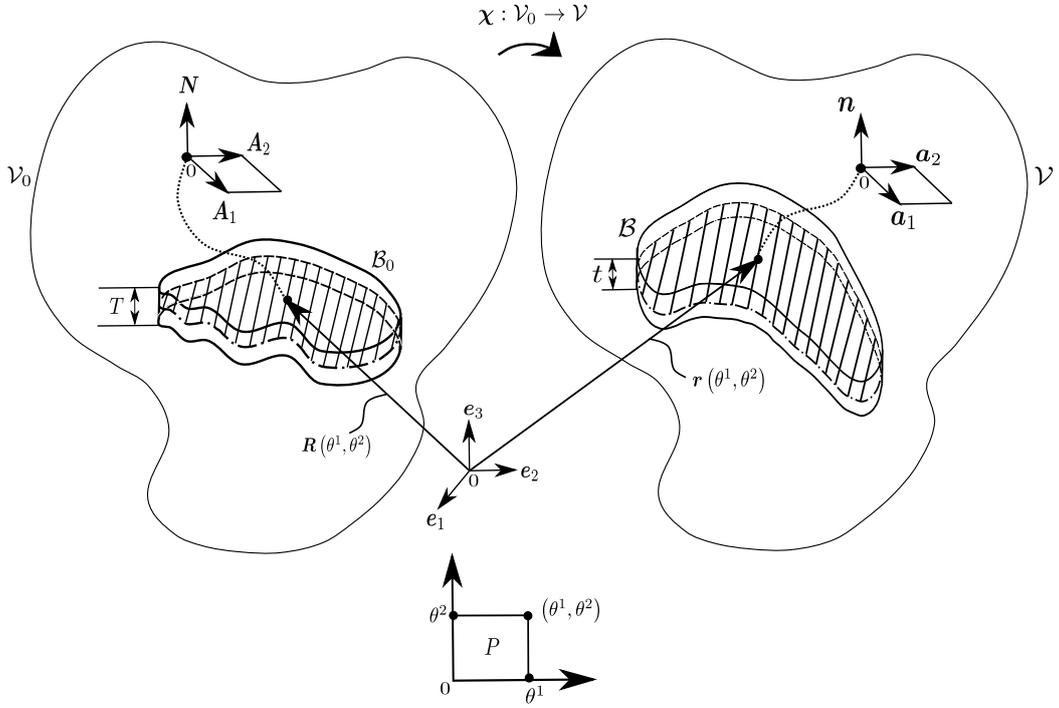}
    \caption{Schematic of a thin shell that occupies regions $\mcal{B}_0$ and $\mcal{B}$ in its reference and current configurations, respectively. The body is embedded in volumes ${\mathcal{V}}_0$ and $\mathcal{V}$ in the two configurations connected by a deformation map $\chi$.
    The two-dimensional parametric coordinate system is denoted by $P$ and the local triads in the two configurations are also shown.}
      \label{fig:shell_definition}
  \end{center}
\end{figure}
The three-dimensional region $\mathcal{B}$ is enclosed within the region $\mathcal{V}$, as schematically shown in Figure \ref{fig:shell_definition}, so that
the surrounding free space is
$ {\mathcal{B}}^{\mathrm{'}} = \mathcal{V} \setminus \mathcal{B} \cup \partial \mathcal{B}.$
 ${\mathcal{V}}_{\mathrm{0}}$ is the referential region corresponding to $ \mathcal{V}$ in ${\mathbb{R}}^3$, such that
$
   {\mathcal{B}}^{'}_{\mathrm{0}}=  {\mathcal{V}}_{\mathrm{0}} \setminus {\mathcal{B}}_{\mathrm{0}} \cup \partial {\mathcal{B}}_{\mathrm{0}}.
$
The deformation gradient $\bm{F}$ is defined by 
\begin{equation}
\bm{F}=\frac{\partial \bm{\chi}}{\partial \bm{X}},
\label{eq:deformation_gradient_1}
\end{equation}
with the Jacobian, $J=\mathrm{det}\bm{F} > 0$, such that
\begin{equation}
    dv=JdV,
    \label{eq:volume_map}
\end{equation}
where $dV$ and $dv$ are the volume elements in the reference and deformed configuration, respectively.
 The right  Cauchy-Green tensor is defined by
\begin{equation}
\bm{C}={\bm{F}}^{\mathrm{T}}\bm{F}. 
\label{eq:right_left_stretch}
\end{equation}  
To facilitate the description of fields exterior to the body. Consider  a fictitious deformation map ${\bm{\chi}}_{\mathrm{F}}$ as 
$
   {\bm{\chi}}_{\mathrm{F}}  ( {\bm{X}}_{\mathrm{F}}): {\mathcal{B}}^{'}_{\mathrm{0}} \rightarrow {\mathcal{B}}^{'}
$
with ${\bm{X}}_{\mathrm{F}} \in {\mathcal{B}}^{'}_{\mathrm{0}}$, satisfying
\begin{equation}
    {\bm{\chi}}_{\mathrm{F}} = {\bm{\chi}}_{\mathrm{B}} \ \text{on} \ \partial \mathcal{B}.
    \label{eq:continuity}
\end{equation}
The boundaries $\partial {\mathcal{V}}_{\mathrm{0}}$ and $\partial \mathcal{V}$ coincide, 
implying
\begin{equation}
    {\bm{x}}_{\mathrm{F}} = {\bm{X}}_{\mathrm{F}} \ \text{on} \ \partial \mathcal{V}.
    \label{eq:ext_bound_chi}
\end{equation}
The magnetostatic problem is governed by the Maxwell's equations involving the spatial magnetic induction $\bm{\bbm{b}}$ and 
magnetic field $\bm{\bbm{h}}$ given by
\begin{eqnarray}
\mathrm{div}  \bm{\bbm{b}}  = 0 \quad \text{and} \quad
\mathrm{curl}  \bm{\bbm{h}}  = \mathbf{0}, \quad \text{in} \quad \mathcal{V},
\label{eq: maxwell b h spatial}
\end{eqnarray}
along with the boundary (or jump) conditions,
\begin{equation}
    \jump{\bbbm{b}} \cdot \mbf{n} = 0 \quad \text{and} \quad \jump{\bbbm{h}} \times \mbf{n} = \mbf{0}, \quad \text{on} \quad \partial \mathcal{B},
    \label{eq: b h spatial bcs}
\end{equation}
where $\jump{\bullet}$ represents jump in a quantity across the boundary with a unit outward normal vector $\mbf{n}$.
Equation \eqref{eq: maxwell b h spatial}$_2$ motivates the introduction of a magnetic scalar potential  $\phi$ such that 
\begin{equation}
\bm{\bbm{h}}=-\mathrm{grad}\phi.
\label{eq:H-Phi_relation}
\end{equation}
Moreover, $\phi$ is continuous across the boundary between the magnetoelastic body and the surrounding space.
The vectors $\bm{\bbm{b}}$ and $\bm{\bbm{h}}$ are related via the well-known constitutive relation, 
\begin{equation}
\bm{\bbm{b}}=\mu_0\bm{\bbm{h}} + \bm{\bbm{m}},
\label{eq:magneticc_constitutive}
\end{equation}
where $\mu_0$ is the constant magnetic permeability of free space and $\bm{\bbm{m}}$ is the spatial magnetisation that vanishes in $\mathcal{B}^{'}$. 
The pull-back transformations of $\bm{\bbm{b}}$, $\bm{\bbm{h}}$, and $\bbbm{m}$ to the reference configuration are given by
\begin{eqnarray}
{\bm{\bbm{B}}} = J \bm{F}^{-1} \bm{\bbm{b}}, \quad \quad 
{\bm{\bbm{H}}} =  \bm{F}^{\mathrm{T}}\bm{\bbm{h}}, \quad \text{and} \quad {\bm{\bbm{M}}}={\bm{F}}^{\mathrm{T}} \bm{\bbm{m}},
\label{transformations}
\end{eqnarray}
thereby allowing one to rewrite the governing Maxwell's equations \eqref{eq: maxwell b h spatial} and the boundary conditions \eqref{eq: b h spatial bcs} in the reference configuration as
\begin{eqnarray}
\mathrm{Div}  \bm{\bbm{B}} 
= 0, \quad \quad 
\mathrm{Curl}  \bbbm{H}
= \mathbf{0},
\end{eqnarray}
and
\begin{equation}
    \jump{\bbbm{B}} \cdot \mbf{N} = 0, \quad \quad \jump{\bbbm{H}} \times \mbf{N} = \mbf{0}, 
    \label{eq: b h referential bcs}
\end{equation}
with $\mbf{N}$ being the outward unit normal to the boundary in the reference configuration.
 Denoting ${\Phi}$ as the referential counterpart of $\phi$, then $\Phi
\left({\bm{X}}\right)=\phi
(\bm{x}) 
\circ \bm{\chi}\left({\bm{X}}\right)
 $ and
\begin{equation}
    {\bm{\bbm{H}}}=- \mathrm{Grad} \Phi.
\end{equation}
 Using the transformations (\ref{transformations}) in the constitutive relation (\ref{eq:magneticc_constitutive}), one obtains
\begin{equation}
J^{-1} \bm{C}{\bm{\bbm{B}}}=\mu_0{\bm{\bbm{H}}}+ {\bm{\bbm{M}}} \ \text{in} \ {\mathcal{V}}_{\mathrm{0}}.
\label{eq:magneticc_constitutive_ref}
\end{equation}
Since, $\bm{\bbm{m}}$ and ${\bm{\bbm{M}}}$ vanish in the vacuum, the constitutive relation simplifies to
\begin{equation}
    {\bm{\bbm{B}}}=\mu_0 J {\bm{C}}^{-1} {\bm{\bbm{H}}} \ \text{in} \ {\mathcal{B}}^{'}_{\mathrm{0}}.
    \label{free_sp_constitutive}
\end{equation}

Coupled magnetoelastic constitutive relations in the body are established  by assuming a
free energy density function per unit reference volume, $\Omega$, that is of the form
$
\Omega=\Omega \left( \bm{F},{{\bm{\bbm{H}}}} \right).
$
Objectivity and isotropy require that the free energy take the form 
\begin{equation}
\Omega = \breve{{\Omega}}\left(\bm{C},\textcolor{black}{{\bm{\bbm{H}}}} \right)=\widetilde{{\Omega}}\left(I_1,I_2,I_3,I_4,I_5,I_6\right),
\label{eq:free_energy2}
\end{equation}
where $I_1, I_2, I_3$ are scalar invariants of $\bm{C}$, that is 
\begin{equation}
I_1 = \mathrm{tr}\bm{C}, \quad
I_2 = \dfrac{1}{2}\left[{\left[\mathrm{tr}\bm{C}\right]}^{2}-\mathrm{tr}\bm{C}^2\right],  \quad \text{and} \quad
I_3 = \mathrm{det}\bm{C}=J^2,
\end{equation}
and the remaining three scalar invariants are given by
\begin{equation}
I_4 = \textcolor{black}{{\bm{\bbm{H}}}}\cdot \textcolor{black}{{\bm{\bbm{H}}}}, \quad
I_5 = \left[\bm{C}\textcolor{black}{{\bm{\bbm{H}}}}\right]\cdot\textcolor{black}{{\bm{\bbm{H}}}}, \quad \text{and} \quad
I_6 = [{\bm{C}}^{2}\textcolor{black}{{\bm{\bbm{H}}}}]\cdot\textcolor{black}{{\bm{\bbm{H}}}}.
\end{equation}
 Incompressibility requires $J \equiv 1$ and the energy density function is further simplified to
\begin{equation}
    \widetilde{{\Omega}}\left(I_1,I_2,I_3,I_4,I_5,I_6\right)=\grave{{\Omega}}\left(I_1,I_2,I_4,I_5,I_6\right).
\end{equation}
The total Piola stress tensor is given for the case of incompressible solid as
\begin{equation}
\bm{P}= \dfrac{\partial {\Omega}}{\partial\bm{F}} - p{\bm{F}}^{-\mathrm{T}},
\label{eq:cauchy_constitutive_general}
\end{equation}
 where $p$ is the  Lagrange multiplier due to the incompressibility constraint. 
The constitutive relation for the magnetic induction is given as
\begin{equation}
    {\bm{\bbm{B}}}=-\dfrac{\partial {\Omega}}{\partial {\bm{\bbm{H}}}}.
    \label{eq:mag_ind_rel_shell}
\end{equation} 

The Maxwell stress tensor outside the body is given by
\begin{equation}
\bm{\sigma}_\mathrm{M}=\frac{1}{\mu_0}\left[\bm{\bbm{b}}\otimes \bm{\bbm{b}}-\frac{1}{2}
\left[\bm{\bbm{b}}\cdot \bm{\bbm{b}}\right] \mathbold{\mathbb{1}} \right],  
\end{equation}
where $\mathbold{\mathbb{1}}$ is the spatial identity tensor.
Using the Piola transform $\bm{P}_{\mathrm{M}}= J \bm{\sigma}_\mathrm{M}\bm{F}^{-\mathrm{T}}$, this can be written in the reference configuration as
\begin{equation}
    {\bm{P}}_{\mathrm{M}}=\mu_0 J \left[
    \left[\bm{F}^{-\mathrm{T}} {\bm{\bbm{H}}}\right] \otimes \left[\bm{F}^{-\mathrm{T}} {\bm{\bbm{H}}}\right] 
    - \frac{1}{2} \underbrace{\left[\bm{F}^{-\mathrm{T}} {\bm{\bbm{H}}}\right]}_{\bbbm{h}} \cdot \left[\bm{F}^{-\mathrm{T}} {\bm{\bbm{H}}}\right] \mathbold{\mathbb{1}}_0 
    \right] \bm{F}^{-\mathrm{T}},
    \label{eq:Maxwell_stress_outside}
\end{equation}
where $ \mathbold{\mathbb{1}}_0 $ is the two-point identity tensor.
\section{Kirchhoff-Love magnetoelastic thin shell}
\subsection{Geometry}
 Consider the magnetoelastic body to be a thin shell. 
 Each point $\bm{X}_{\mathrm{B}} \in {\mathcal{B}}_{\mathrm{0}}$ is mapped from the parametric domain defined by the coordinate system $\{ \theta^1, \theta^2, \eta\}$.
 The Kirchhoff-Love hypothesis states that for thin shell structures, lines perpendicular to the mid-surface of the shell remain straight and perpendicular to the mid-surface after deformation \citep[see e.g.][]{Niordson1985}.
 Hence, assuming the shell has a thickness $T\left(\theta^\alpha\right)$ in the reference configuration, the point $\bm{X}_{\mathrm{B}}$  can be defined using a point on the mid-surface ${S}_{\mathrm{m}}$ of the shell, ${\bm{R}} \in {S}_{\mathrm{m}}$, and the associated unit normal vector ${\bm{N}}$ as
 \begin{equation}
\bm{X}_{\mathrm{B}}  = {\bm{R}} + \eta {\bm{N}},
\label{eq: r x n relation}
 \end{equation}
 where $\eta \in [-T/{2}, T/{2}]$. 
The points on the mid-surface in the deformed configuration denoted by $\bm{r}$. 
The mid-surface point in the deformed configuration corresponds to the mid-surface point in the reference configuration after a motion as shown in Figure \ref{fig:shell_definition}, and can be expressed as
\begin{equation}
\bm{r} = \bm{R} + \mathbf u,
\label{eq:reference_to_deformed}
\end{equation}
where $\bm{u}$ denotes the mid-surface displacement. A point $\bm{x}_{\mathrm{B}} \in \mathcal{B}$ can therefore be expressed as 
\begin{equation}
\bm{x}_{\mathrm{B}}=\bm{r}+\eta \bm{d}, 
\label{eq:def_map_body}
\end{equation} 
where $\bm{d}=\lambda \bm{n}$, and $\lambda$ is the through thickness stretch for a finitely deformed shell defined by
\begin{equation}
\lambda= \frac{t}{T},
\end{equation}
where $t(\theta^{\alpha})$ is the shell thickness after deformation.  
Further,  in ${\mathcal{B}}_0$, one can assume a form for the magnetic potential as 
\begin{eqnarray}
\Phi \left(\eta, \theta^\alpha \right)
= {\Phi}_0\left(\theta^\alpha\right) +  \eta {\Phi}_1\left(\theta^\alpha\right)+ \frac{\eta^2}{2}  {\Phi}_2\left(\theta^\alpha\right)+\mathcal{O}\left(\eta^3\right),
\end{eqnarray}
where ${\Phi}_0=\Phi \left(0, \theta^\alpha \right)$. Furthermore, assume
\begin{equation}
 {\Phi}_1\left(\theta^\alpha\right)=\frac{\Phi\left(T/2, \theta^\alpha \right)-\Phi\left(-T/2, \theta^\alpha \right)}{T} , 
 \label{eq:potential_thin_shell_assumption}
\end{equation}
implying a linear variation of the magnetic potential along the thickness of the thin shell. Therefore, the higher-order terms vanish, allowing one to express the Kirchhoff-Love assumption as
\begin{equation}
 \Phi\left(\eta, \theta^\alpha \right)={\Phi}_0\left(\theta^\alpha\right) +  \eta {\Phi}_1\left(\theta^\alpha\right), 
 \label{eq:kirchoff_love_equivalent}
\end{equation}
which is similar in form to Equation~\eqref{eq: r x n relation}.

 Table \ref{demo-table1} provides a list of  surface parameters used to describe the geometry of the shell and Table \ref{demo-table2} presents the surface and volume elements of the shell. The expressions and  associated derivations are elaborated on in Appendix A.
The boundaries, $\partial{\mathcal{B}}_{\mathrm{0}}$ and $\partial \mathcal{B}$, can be written as
$ \partial{\mathcal{B}}_{\mathrm{0}}= S_\mathrm{t} \cup   S_\mathrm{b}  \cup   S_\ell$, and $\partial \mathcal{B}= s_\mathrm{t} \cup  s_\mathrm{b} \cup  s_\ell$,
where the subscripts, t, b, and $\ell$, represents the top, bottom, and  lateral surfaces in the two configurations, and the top surface is the side of the boundary that is reached along the unit outward normal vector.
\begin{table}
\small
\centering
\caption{\label{demo-table1} The parameters used to describe the geometry of the thin shell in reference and deformed configurations.}
\resizebox{\textwidth}{!}{
 \begin{tabular}{||l l l||} 
 \hline
 Surface Parameters & Reference Configuration & Deformed Configuration \\ [0.4ex] 
 \hline\hline
 Covariant basis vectors at the mid-surface  & $\bm{A}_\alpha$ & $\bm{a}_\alpha$ \\
 Covariant metric tensor at the mid-surface &$ A_{\alpha \beta}$ & 
 $a_{\alpha \beta}$ \\  
  Determinant of the covariant metric tensor at the mid-surface &$ A $ & 
 $a$ \\ 
 Contravariant basis vectors at a mid-surface  & $\bm{A}^\alpha$ & $\bm{a}^\alpha$ \\
 Contravariant metric tensor at the mid-surface & $A^{\alpha \beta}$ & $a^{\alpha \beta}$  \\
  Determinant of the contravariant metric tensor at the mid-surface & $A^{-1}$ & $a^{-1}$  \\
 Christoffel symbol at the mid-surface & ${\Gamma}^\alpha_{\beta \gamma}$ & ${\gamma}^\alpha_{\beta \gamma}$\\
 Parametric derivative of the metric at the mid-surface & ${A}_{,\zeta}
 $ & ${a}_{,\zeta}
 $ \\
  Normal  at the mid-surface & $\bm{N}$ & $\bm{n}$ \\
 Tangent  on the bounding curve of the  mid-surface  & $\bm{\tau}$ & $--$ \\
 In-plane normal on the bounding curve of the  mid-surface & $\bm{\nu}$ & $--$ \\
  Projection tensor at the mid-surface & $\bm{I}$ & $\bm{i}$ \\
 Curvature tensor at the mid-surface & $\bm{K}$ & $\bm{\kappa}$ \\
 Mean curvature at the mid-surface & $H$ & $h$ \\
 Gaussian curvature at the mid-surface & $K$ & $\kappa$ \\
 Covariant basis vectors at a shell-point & $\bm{G}_\alpha=\bm{M} \bm{A}_\alpha \ \text{and} \ \bm{M}=\bm{I}-\eta \bm{K}$ & $\bm{g}_\alpha=\bm{\mu} \bm{a}_\alpha \ \text{and} \ \bm{\mu}=\bm{i}-\eta \lambda\bm{\kappa}$  \\
 Covariant metric tensor at a shell-point & $G_{\alpha \beta}$ & $g_{\alpha \beta}$\\
 Contravariant basis vectors at a shell-point & ${\bm{G}}^{\alpha}={\bm{M}}^{-\mathrm{T}} {\bm{A}}^{\alpha}$ & ${\bm{g}}^{\alpha}={\bm{\mu}}^{-\mathrm{T}} {\bm{a}}^{\alpha}$  \\
 Contravariant metric tensor at a shell-point & $G^{\alpha \beta}$  & $g^{\alpha \beta}$   \\
 Tangent  at a shell-point on the lateral surface & $\bm{\tau}_\ell=\bm{M}\bm{\tau}/ c \ \text{and} \ c=\norm{\bm{M}\bm{\tau}}$ & $--$ \\
 In-plane normal  at a shell-point on the lateral surface & $\bm{\nu}_\ell={c}^{-1}\left[\bm{I}+\eta\left[\bm{K}-2H\bm{I}\right]\right]\bm{\nu}$ & $--$ \\
  [1ex] 
 \hline
 \end{tabular}}
\end{table}
\begin{table}
\centering
\caption{\label{demo-table2} Description of the surface and volume elements used for integration.}
\resizebox{\textwidth}{!}{
 \begin{tabular}{||l l l||} 
 \hline
 Volume/Surface elements  & Reference Configuration & Deformed Configuration  \\ [0.5ex] 
 \hline\hline
 Elemental area for the convected coordinates & $dP$ & $dP$ \\
 Elemental area at the curved mid-surface & $dS_\mathrm{m}$ & $ds_\mathrm{m}$ \\ 
 Elemental area at a shell-point & $dS=M d{S}_\mathrm{m}\ \text{and} \ M=\mathrm{det} \bm{M}$ & $ds=\mu {\hat{a}}^{1/2} dS_\mathrm{m}, \ \mu=\mathrm{det} \bm{\mu}, \ \text{and} \ \hat{a}=a/A$ \\
 Elemental area at the top surface & $dS_\mathrm{t}$  & $ds_\mathrm{t}$  \\
 Elemental area at the bottom surface & $dS_\mathrm{b}$  & $ds_\mathrm{b}$  \\
 Elemental area on the lateral surface & $dS_\ell=c \, dl \, d\eta$  & $--$  \\
  Elemental volume at a shell-point & $dV=dS \, d\eta$ & $--$ \\
 [1ex] 
 \hline
 \end{tabular}
 }
\end{table}

The incorporation of the variation of the through-thickness stretch and the deformed normal in obtaining the reduced-order model for the soft thin magnetoelastic shell requires the evaluation of the integral: $\int\limits_{ P}\left[{A}^{1/2}{T}^\alpha\right]_{, \alpha}dP$ for an arbitrary $T^\alpha \left(\theta^\beta\right)$, as discussed  in Sections 5.1 and 5.3. This integral can be expressed as:
\begin{equation}
\int\limits_{ P}\left[{A}^{1/2}{T}^\alpha\right]_{, \alpha}dP=\int\limits_{\mathcal{C}_\mathrm{m}}{T}^\alpha {\nu}_\alpha dl,
   \label{eq:dom_to_bound}
\end{equation}
where $\mathcal{C}_\mathrm{m}$ represents the boundary of the curved mid-surface $S_\mathrm{m}$. This is elaborated upon further in Appendix B. 
\subsection{Kinematics}
The deformation gradient and its inverse for a shell-point 
can be written so as to separate the thickness variable from the surface parameters as
\begin{eqnarray}
\bm{F}
&=&\left[{F}^{ \alpha}_{0_\beta}
+\eta {F}^{  \alpha}_{1_\beta}
+ \eta^2 {F}^{ \alpha}_{2_\beta}\right] 
\bm{a}_{\alpha} \otimes {\bm{A}}^{\beta} +\lambda \bm{n}\otimes\bm{N}
 + \bm{\mathcal{O}}(\eta^3),
\label{def_gradient}
\end{eqnarray}
with
\begin{eqnarray}
{F}^{ \alpha}_{0_\beta}=\delta^\alpha_\beta,\quad
{F}^{ \alpha}_{1_\beta}=-\lambda {b}_\beta^{ \ \alpha}+{B}_\beta^{ \ \alpha}, \quad \text{and} \quad 
{F}^{  \alpha}_{2_\beta}={B}_\delta^{ \ \alpha}{B}_\beta^{ \ \delta}-\lambda {b}_\delta^{ \ \alpha} {B}_\beta^{ \ \delta},
\end{eqnarray}
and
\begin{eqnarray}
\bm{F}^{-1}
&=&\left[{F}^{-1 \alpha}_{0_\beta}
+
\eta {F}^{-1  \alpha}_{1_\beta}
+ 
\eta^2 {F}^{-1  \alpha}_{2_\beta}\right] {\bm{A}}_{\alpha} \otimes \bm{a}^{\beta} 
+\frac{1}{\lambda} \bm{N}\otimes \bm{n}+ \bm{\mathcal{O}}(\eta^3) ,
\label{inv_def_gradient}
\end{eqnarray}
with
\begin{equation}
{F}^{-1 \alpha}_{0_\beta}=\delta^\alpha_\beta,\quad
{F}^{-1  \alpha}_{1_\beta}=\lambda {b}_\beta^{ \ \alpha}-{B}_\beta^{ \ \alpha}, \quad \text{and} \quad
{F}^{-1  \alpha}_{2_\beta}=\lambda^2{b}_\delta^{ \ \alpha}{b}_\beta^{ \ \delta}-\lambda {B}_\delta^{ \ \alpha} {b}_\beta^{ \ \delta}.
\end{equation}
Here ${B}_\beta^{ \ \alpha}$ and ${b}_\beta^{ \ \alpha}$ are the components of the curvature tensors $\bm{K}$ and $\bm{\kappa}$, respectively.
Further, using the relation, $\bm{a}_\alpha \cdot \bm{n}=0$, the right Cauchy-Green deformation tensor  can be written as 
\begin{equation}
\bm{C}
=\left[{C}_{0_{\alpha\beta}} 
+ 
\eta {C}_{1_{\alpha\beta}}
+ 
\eta^2 {C}_{2_{\alpha\beta}}\right] {\bm{A}}^{\alpha}\otimes {\bm{A}}^{\beta} + \lambda^2 \bm{N}\otimes \bm{N}+ \bm{\mathcal{O}}(\eta^3),
\end{equation}
where
\begin{eqnarray}
{C}_{0_{\alpha\beta}}&=&a_{\alpha \beta},\nonumber \\
{C}_{1_{\alpha\beta}}&=&-\lambda {b}_\beta^{ \ \gamma}a_{\gamma \alpha}-\lambda {b}_\alpha^{ \ \gamma}a_{\gamma \beta} +{B}_\alpha^{ \ \gamma}a_{\gamma \beta}+{B}_\beta^{ \ \gamma}a_{\gamma \alpha}, \nonumber\\
{C}_{2_{\alpha\beta}}&=&{B}_\alpha^{ \ \gamma}{B}_\beta^{ \ \delta}a_{\delta \gamma }
+\lambda^2 {b}_\alpha^{ \ \delta}{b}_\beta^{ \ \gamma}a_{\delta \gamma }
+{B}_\gamma^{ \ \delta}{B}_\alpha^{ \ \gamma}a_{\delta \beta }  
+{B}_\gamma^{ \ \delta}{B}_\beta^{ \ \gamma}a_{\delta \alpha }
-\lambda {b}_\gamma^{ \ \delta} {B}_\beta^{ \ \gamma}a_{\delta \alpha } 
-\lambda {b}_\gamma^{ \ \delta} {B}_\alpha^{ \ \gamma}a_{\delta \beta } 
\nonumber \\
&\textcolor{white}{=}&- \lambda {B}_\beta^{ \ \gamma}{b}_\alpha^{ \ \delta} a_{\delta \gamma } 
-\lambda {B}_\alpha^{ \ \gamma}{b}_\beta^{ \ \delta} a_{ \gamma \delta}.
\end{eqnarray}
The mid-surface right Cauchy-Green tensor is defined by
\begin{equation}
    \bm{C}_\mathrm{m}=\bm{C}\Big|_{\eta=0}={C}_{0_{\alpha\beta}} {\bm{A}}^{\alpha}\otimes {\bm{A}}^{\beta} +\lambda^2 \bm{N}\otimes \bm{N},
\end{equation}
with
\begin{equation}
    J_0=\mathrm{det}\bm{C}_\mathrm{m}
    =\mathrm{det}\left[{C_{\mathrm{m}_j}}^{  i}\right]
    =\mathrm{det}\left[{C}_{\mathrm{m}_{ j k}}{A}^{ki }\right]
    =\frac{ \mathrm{det}\left[{C}_{\mathrm{m}_{ ij}}\right]}{\mathrm{det}\left[{A}_{ij }\right]}=\frac{ \mathrm{det}\left[a_{\alpha \beta}\right] \lambda^2}{\mathrm{det}\left[{A}_{\alpha \beta}\right]}
    =\frac{ a \lambda^2}{A},
\end{equation}
where ${A}_{i j} = {\bm{A}}_{i} \cdot {\bm{A}}_{j}$ and ${A}^{i j} = {\bm{A}}^{i} \cdot {\bm{A}}^{j}$ are the components of the three-dimensional covariant and contravariant metric tensors  on the mid-surface, respectively,  with $ {\bm{A}}_{3}= 
    {\bm{A}}^{3}
    ={\bm{N}}$.
The incompressibility constraint,
$J=1$, 
implies that
\begin{equation}
    \lambda={\hat{a}}^{-1/2} ,
    \label{eq:through_thickness}
\end{equation}
where the surface stretch $\hat{a}$ is defined in Table \ref{demo-table2}.
\subsection{Divergence of the total stress tensor and magnetic induction vector }
The divergence of the total Piola stress tensor, as well as the magnetic induction vector, enters the governing equations for the Kirchhoff-Love magnetoelastic thin shell arising from the variational formulation involving the mechanical deformation and an independent field representing the magnetic component, that is,  the magnetic field vector. 
The divergence of the total stress tensor can be expressed as
\begin{eqnarray}
   \mathrm{Div}\bm{P}
     =\bm{A}_0+\eta \bm{A}_1 + \eta^2 \bm{A}_2 +\bm{\mathcal{O}}(\eta^3),
  \label{eq:div_total_stress}
\end{eqnarray}
where
\begin{eqnarray}
  \bm{A}_0&=&{\bm{P}_0}_{, \alpha}  {\bm{A}}^\alpha + \bm{P}_1  \bm{N}, \nonumber \\
  \bm{A}_1&=&{B}_\delta^{ \ \alpha} {\bm{P}_0}_{, \alpha}  {\bm{A}}^\delta+{\bm{P}_1}_{, \alpha} {\bm{A}}^\alpha,
  +\bm{P}_2 \bm{N}, \nonumber \\
  \bm{A}_2 &=& {B}_\zeta^{ \ \alpha}{B}_\delta^{ \ \zeta}{\bm{P}_0}_{, \alpha}  {\bm{A}}^\delta + {B}_\delta^{ \ \alpha} {\bm{P}_1}_{, \alpha}  {\bm{A}}^\delta + \frac{1}{2}{\bm{P}_2}_{, \alpha} {\bm{A}}^\alpha 
 + \frac{1}{2}\bm{P}_3  \bm{N},
\end{eqnarray}
with 
$\bm{P}=\bm{P}(\eta, \theta^\alpha)=\bm{P}_0+\eta \bm{P}_1 + \dfrac{\eta^2}{2}\bm{P}_2  
+\bm{\mathcal{O}}(\eta^3)$.
Similarly, for the magnetic induction vector at a shell-point,
\begin{eqnarray}
   \mathrm{Div}\bm{\bbm{B}}
   ={B}_0+\eta {B}_1 + \eta^2 {B}_2 +\mathcal{O}(\eta^3),
   \label{eq:div_of_mag_ind}
\end{eqnarray}
where
\begin{eqnarray}
  {B}_0&=&{{\bm{\bbm{B}}}_0}_{, \alpha} \cdot {\bm{A}}^\alpha + {\bm{\bbm{B}}}_1 \cdot \bm{N}, \nonumber \\
  {B}_1&=&{B}_\delta^{ \ \alpha} {{\bm{\bbm{B}}}_0}_{, \alpha} \cdot {\bm{A}}^\delta+
  {{\bm{\bbm{B}}}_1}_{, \alpha} \cdot {\bm{A}}^\alpha
  + {\bm{\bbm{B}}}_2 \cdot \bm{N}, \nonumber \\
  {B}_2&=& {B}_\zeta^{ \ \alpha}{B}_\delta^{ \ \zeta}{{\bm{\bbm{B}}}_0}_{, \alpha} \cdot {\bm{A}}^\delta + {B}_\delta^{ \ \alpha} {{\bm{\bbm{B}}}_1}_{, \alpha} \cdot {\bm{A}}^\delta + \frac{1}{2}
  {{\bm{\bbm{B}}}_2}_{, \alpha}\cdot {\bm{A}}^\alpha 
 + \frac{1}{2}{\bm{\bbm{B}}}_3\cdot \bm{N},
\end{eqnarray}
with ${\bm{\bbm{B}}}=
    \bm{\bbm{B}} (\eta, \theta^\alpha)
    ={\bm{\bbm{B}}}_0+\eta {\bm{\bbm{B}}}_1+ \dfrac{\eta^2}{2}{\bm{\bbm{B}}}_2 +\bm{\mathcal{O}}(\eta^3)$.
The total Piola stress tensor and  magnetic induction vector at the top and bottom boundaries are obtained by setting  $\eta= \pm   T/2$ in their respective through-thickness expansions.
\subsection{Interface condition on magnetic field}
The continuity of magnetic field components projected onto the tangent space across the boundaries of the thin shell and the surrounding space is implied by Equation~\eqref{eq: b h referential bcs}$_2$. Thereby, at the interfaces, by equating these components and using Equation~\eqref{def_grad_deriv_3}, the resulting expression can be written as
\begin{equation}
-\mathrm{Grad}\Phi \cdot \bm{A}_\alpha + \eta \mathrm{Grad}\Phi B_\alpha^{\ \gamma} \cdot \bm{A}_\gamma  +{\Phi_0}_{, \alpha}  +\eta \left[{\Phi_0}_{, \beta}B_\alpha^{\ \beta}  + {\Phi_1}_{, \alpha} \right]
 +\mathcal{O}(\eta^2)=0,
  \label{eq:ext_bound_mag_field}
 \end{equation}
 where $\mathrm{Grad}\Phi$ is evaluated in the free space. 
 This imposes a constraint on the potential at the top, bottom, and lateral surfaces of the shell. 
 Note, to the continuity of the potential across the boundaries of the thin shell with the surrounding space, this constraint is explicitly imposed and is not obtained from the modified variational setting while deriving the reduced-order theory, as discussed in Section 5.
\section{Variational formulation in three dimensions}
 Defining   $\grave{\bm{\chi}}=\{\bm{\chi},  \Phi, p\}$ as the generalised set of the solution variables, the total potential energy of the system is written as \citep{dorfmann2014}:
\begin{eqnarray}
    \Pi[\grave{\bm{\chi}}]&=&\int\limits_{{\mathcal{B}}_0}\Omega\left( \bm{F},\bm{\bbm{H}} \right)dV-
    \int\limits_{\mathcal{B}_0}p[J-1]dV-
    \frac{\mu_0}{2}\int\limits_{\mathcal{B}^{'}}\bm{\bbm{h}}\cdot\bm{\bbm{h}}dv
    -\int\limits_{\partial \mathcal{V}}\phi \bm{\bbm{b}}_\mathrm{e} \cdot \bm{n}^{'}ds \nonumber \\
    &\textcolor{white}{=}&-\int\limits_{\mathcal{B}_0}\bm{\mathfrak{B}}\cdot \bm{\chi}dV
    -\int\limits_{ {\mathcal{C}}_\mathrm{m} \setminus {\mathcal{C}}^{\mathrm{u}}_{\mathrm{m}}}
    \bm{t}_\ell\cdot\bm{\chi}dl
    -\int\limits_{ s_\mathrm{t}}\bm{p}_\mathrm{t}\cdot \bm{\chi}ds_\mathrm{t}
     -\int\limits_{ s_\mathrm{b}}\bm{p}_\mathrm{b} \cdot \bm{\chi}ds_\mathrm{b}\ .
     \label{eq:energy}
\end{eqnarray}
The external spatial magnetic induction is denoted as $\bm{\bbm{b}}_\mathrm{e}$. 
Its  normal component is prescribed on $\partial \mathcal{V}$ and  its counterpart in the reference configuration is denoted as ${\bm{\bbm{B}}}_\mathrm{e}$.
The fourth term in Equation~\eqref{eq:energy} representing the work done by the external magnetic induction is expressed in the current configuration, and  using Equation~\eqref{eq:ext_bound_chi} can
be rewritten in the reference configuration as
\begin{equation}
    \int\limits_{\partial \mathcal{V}}\phi \bm{\bbm{b}}_e \cdot \bm{n}^{'}ds=
    \int\limits_{\partial {\mathcal{V}}_0}\Phi {\bm{\bbm{B}}}_e \cdot {\bm{N}}^{'}dS \ ,
\end{equation}
with the associated unit normals on the outer boundary of the free space,  denoted by  ${\bm{N}}'$ and $\bm{n}^{'}$ in the reference and deformed configurations, respectively. 
The body force field per unit reference volume is $\bm{\mathfrak{B}}$ while $\bm{t}_\ell$ is the applied traction 
at  ${C}_\mathrm{m}$. Also, ${C}_\mathrm{m}^\mathrm{u}$ are the parts of the boundary where displacements are specified. ${p}_\mathrm{t}\left(\theta^\alpha\right)$ and ${p}_\mathrm{b}\left(\theta^\alpha\right)$ are the magnitudes of external pressure at the top and bottom surfaces of the shell, respectively, such that 

\begin{equation}
    \bm{p}_\mathrm{t}=-p_\mathrm{t} \bm{n}
    , \quad \text{and} \quad \bm{p}_\mathrm{b}=p_\mathrm{b}\bm{n}
    \ , \label{eq:press_term_forms}
\end{equation}
with $\bm{n}$
is the  mid-surface unit normal 
in the current configuration.
Let the set $\delta\grave{\bm{\chi}}=\{\delta\bm{\chi},  \delta \Phi, \delta p\}$. In the subsequent calculations, refer to Appendix \ref{app: variation of relevant quantities}  for details of
the variation of key variables.  From Equations ~\eqref{eq:volume_map} and \eqref{transformations}, the first variation of the total energy is given by
\begin{eqnarray}
    \delta \Pi[\grave{\bm{\chi}}, \delta \grave{\bm{\chi}}]&=& \delta \left[\int\limits_{{\mathcal{B}}_0}\Omega\left( \bm{F},{{\bm{\bbm{H}}}} \right)dV\right]
    -\int\limits_{{\mathcal{B}}_0}p\delta JdV
    -\int\limits_{{\mathcal{B}}_0}\delta p [J-1]dV
    -\frac{\mu_0}{2}\delta\left[\int\limits_{{\mathcal{B}}^{'}_0}\left[\bm{F}^{-T}\bm{\bbm{H}}\right]\cdot \left[\bm{F}^{-T}\bm{\bbm{H}}\right] J dV\right]\nonumber \\
    &\textcolor{white}{=}&-\delta\left[\int\limits_{\partial {\mathcal{V}}_0}\Phi {\bm{\bbm{B}}}_\mathrm{e} \cdot {\bm{N}}^{'}dS\right]
    -\int\limits_{{\mathcal{B}}_0}\bm{\mathfrak{B}}\cdot \delta\bm{\chi}dV 
     -\int\limits_{ {\mathcal{C}}_\mathrm{m} \setminus {\mathcal{C}}_\mathrm{m}^\mathrm{u}}\bm{t}_\ell \cdot\delta\bm{\chi}dl
    -\int\limits_{ s_\mathrm{t}}\bm{p}_\mathrm{t}\cdot \delta\bm{\chi}ds_\mathrm{t}
     -\int\limits_{ s_\mathrm{b}}\bm{p}_\mathrm{b}\cdot  \delta\bm{\chi}ds_\mathrm{b}\ .
     \label{eq:var_2}
\end{eqnarray}
The Euler--Lagrange equations are obtained by setting $\delta \Pi=0$.
\begin{itemize}
    \item The first and second terms in Equation \eqref{eq:var_2} can be combined  as
\begin{align}
   \delta \left[\int\limits_{{\mathcal{B}}_0}\Omega\left( \bm{F},\bm{\bbm{H}} \right)dV\right]
    -\int\limits_{{\mathcal{B}}_0}p\delta JdV &= \int\limits_{{\mathcal{B}}_0}\left[\frac{\partial\Omega}{\partial \bm{F}}:\delta\bm{F}+\frac{\partial\Omega}{\partial \bm{\bbm{H}}}\cdot \delta\bm{\bbm{H}}\right]dV-
    \int\limits_{{\mathcal{B}}_0}p J \bm{F}^{-\mathrm{T}} : \delta \bm{F}dV \ , \\
\intertext{and taking into account the incompressibility condition ($J=1$), and using Equations ~\eqref{eq:cauchy_constitutive_general}, \eqref{eq:mag_ind_rel_shell}, and  \eqref{eq:var_mag_field}, the  expression reduces to}
    &= \int\limits_{{\mathcal{B}}_0}\bm{P}:\delta\bm{F}dV+\int\limits_{{\mathcal{B}}_0}\bm{\bbm{B}}\cdot \frac{\partial \delta \Phi}{\partial \bm{X}}dV \ .
\end{align}
On an application of  the divergence theorem, 
\begin{eqnarray}
 &-& \int\limits_{{\mathcal{B}}_0}\mathrm{Div}\bm{P}\cdot\delta\bm{\chi}dV
 + \int\limits_{ S_\mathrm{t}}\bm{P}{\bm{N}}
 \cdot\delta\bm{\chi}dS_\mathrm{t}
 -\int\limits_{ S_\mathrm{b}}\bm{P}{\bm{N}}
 \cdot\delta\bm{\chi}dS_\mathrm{b} 
 +\int\limits_{ S_\ell}\bm{P}{\bm{\nu}}_\ell\cdot\delta\bm{\chi}dS_\ell 
 \nonumber \\
 &-& \int\limits_{{\mathcal{B}}_0}\mathrm{Div}\bm{\bbm{B}}\delta \Phi dV
 +\int\limits_{ S_\mathrm{t}}\bm{\bbm{B}}\cdot {\bm{N}}
 \delta \Phi dS_\mathrm{t} 
 - \int\limits_{ S_\mathrm{b}}\bm{\bbm{B}}\cdot {\bm{N}}
 \delta \Phi dS_\mathrm{b} 
 +\int\limits_{ S_\ell}\bm{\bbm{B}}\cdot {\bm{\nu}}_\ell \delta \Phi dS_\ell \ .
 \label{eq:var_inside_body}
\end{eqnarray}

\item The fourth term in Equation \eqref{eq:var_2} can be written as
\begin{align}
 -\frac{\mu_0}{2}\delta\left[\int\limits_{{\mathcal{B}}_0^{'}}\left[\bm{F}^{-\mathrm{T}}\bm{\bbm{H}}\right]\cdot \left[\bm{F}^{-\mathrm{T}} \bm{\bbm{H}}\right] J dV \right]
   &=-\frac{\mu_0}{2}\int\limits_{{\mathcal{B}}_0^{'}}J\bm{F}^{-\mathrm{T}}:\delta\bm{F}\left[\bm{F}^{-\mathrm{T}}\bm{\bbm{H}}\right]\cdot\left[\bm{F}^{-\mathrm{T}}\bm{\bbm{H}}\right]dV \nonumber \\
   &\textcolor{white}{=} +\mu_0\int\limits_{{\mathcal{B}}_0^{'}}J\left[\bm{F}^{-\mathrm{T}}{\delta\bm{F}}^{\mathrm{T}}\left[\bm{F}^{-\mathrm{T}}\bm{\bbm{H}}\right]\right]\cdot\left[\bm{F}^{-\mathrm{T}}\bm{\bbm{H}}\right]dV \nonumber \\
   &\textcolor{white}{=} -\mu_0\int\limits_{{\mathcal{B}}_0^{'}}J\left[\bm{F}^{-\mathrm{T}}\bm{\bbm{H}}\right]\cdot\left[\bm{F}^{-\mathrm{T}}\frac{\partial \delta \Phi}{\partial \bm{X}}\right]dV \ , \\
\intertext{and from Equations ~\eqref{eq:Maxwell_stress_outside} and \eqref{free_sp_constitutive},} 
 &= \int\limits_{{\mathcal{B}}_0^{'}}\left[\bm{P}_\mathrm{M}:\delta\bm{F}+\bm{\bbm{B}}\cdot \frac{\partial \delta \Phi}{\partial \bm{X}}\right]dV \ .
\end{align}
Applying the divergence theorem, one obtains
\begin{eqnarray}
 &\textcolor{white}{=}&\int\limits_{{\mathcal{B}}_0^{'}}\left[\bm{P}_\mathrm{M}:\delta\bm{F}+{\bm{\bbm{B}}}^{'}\cdot \frac{\partial \delta \Phi}{\partial \bm{X}}\right]dV
 \nonumber \\
 &=&
 -\int\limits_{{\mathcal{B}}_0^{'}}\mathrm{Div}\bm{P}_\mathrm{M}\cdot\delta\bm{\chi}dV
 +\int\limits_{\partial {\mathcal{V}}_0}\bm{P}_\mathrm{M} {\bm{N}}^{'}\cdot\delta\bm{\chi}dS
 -\int\limits_{ S_\mathrm{t}}{\bm{P}_\mathrm{M}}_{\mathrm{t}} {\bm{N}}
 \cdot\delta\bm{\chi}dS_\mathrm{t}
 +\int\limits_{ S_\mathrm{b}}{\bm{P}_\mathrm{M}}_{\mathrm{b}}{\bm{N}}
 \cdot\delta\bm{\chi} dS_\mathrm{b} 
 -\int\limits_{ S_\ell}{\bm{P}_\mathrm{M}}_\ell {\bm{\nu}}_\ell\cdot\delta\bm{\chi}dS_\ell \nonumber \\
&\textcolor{white}{=}&
 -\int\limits_{{\mathcal{B}}_0^{'}}\mathrm{Div}{\bm{\bbm{B}}}^{'}\delta \Phi dV+\int\limits_{ \partial {\mathcal{V}}_0} {\bm{\bbm{B}}}^{'}\cdot {\bm{N}}^{'}\delta \Phi dS
  -\int\limits_{ S_\mathrm{t}}{\bm{\bbm{B}}}_{\mathrm{t}}^{'}\cdot {\bm{N}}
 \delta \Phi dS_\mathrm{t} 
 +\int\limits
 _{ S_\mathrm{b}}
 {\bm{\bbm{B}}}_{\mathrm{b}}^{'}\cdot {\bm{N}}
 \delta \Phi dS_\mathrm{b} 
  - \int\limits_{ S_\ell} {\bm{\bbm{B}}}_\ell^{'}\cdot  {\bm{\nu}}_\ell\delta \Phi dS_\ell \ .
  \nonumber \\
  \label{eq:var_outside_body}
\end{eqnarray}
From Equation \eqref{eq:ext_bound_chi} it follows that the second term is zero.
The exterior magnetic induction is denoted as ${\bm{\bbm{B}}}^{'}$. Further, at the top and bottom boundaries, ${\bm{\bbm{B}}}^{'}$ is denoted as  ${\bm{\bbm{B}}}_{\mathrm{t}}^{'}$ and ${\bm{\bbm{B}}}_{\mathrm{b}}^{'}$, respectively.
Similarly, the Maxwell stress tensors at the top and bottom surfaces of the shell 
are given by ${\bm{P}_\mathrm{M}}_{\mathrm{t}}$ and ${\bm{P}_\mathrm{M}}_{\mathrm{b}}$, respectively.
For the lateral surface, the expressions for ${\bm{P}_\mathrm{M}}_{\ell}$ and ${\bm{\bbm{B}}}_{\ell}^{'}$ are as follows:
\begin{equation}
{\bm{P}_\mathrm{M}}_{\ell}={\bm{P}_\mathrm{M}}_{0} + \eta {\bm{P}_\mathrm{M}}_1
   +\bm{\mathcal{O}}(\eta^2) \quad \text{and} \quad
{\bm{\bbm{B}}}_{\ell}^{'}={\bm{\bbm{B}}}_{0}^{'}+\eta {\bm{\bbm{B}}}_{1}^{'}
     +\bm{\mathcal{O}}(\eta^2). 
\label{eq:maxwell_mag_ind_lateral}
\end{equation}
\item Since, ${\bm{\bbm{B}}}_\mathrm{e}$ is the applied magnetic induction, the fifth term in Equation \eqref{eq:var_2} can be written as
\begin{eqnarray}
    \delta\left[\int\limits_{\partial {\mathcal{V}}_0} \Phi {\bm{\bbm{B}}}_\mathrm{e} \cdot {\bm{N}}^{'}dS\right]&=&
    \int\limits_{\partial {\mathcal{V}}_0}\delta \Phi 
    {\bm{\bbm{B}}}_\mathrm{e} \cdot {\bm{N}}^{'}dS \ . 
      \label{eq:mag_load}
\end{eqnarray}
\end{itemize}
The remaining terms in Equation  \eqref{eq:var_2}, that is, the virtual work done by the dead load traction, body force and pressures are dealt with in Section 5.3, where their contributions to a modified variational form for a Kirchhoff-Love thin shell  are discussed.
\section{Two dimensional variational formulation for magnetoelastic shells}
The following discussions outline the key steps involved in deriving the Kirchhoff-Love shell equations.
It is important to note that the derived equations can achieve accuracy up to the linear order of the through-thickness parameter. 
The generalised set of solution variables is now extended to $\widetilde{\bm{\chi}}=\{\bm{r}, \bm{\chi}_\mathrm{F}, {\Phi}_0, {\Phi}^{'}, p_0\}$, and define
$\delta \widetilde{\bm{\chi}}=\{\delta\bm{r}, \delta\bm{\chi}_\mathrm{F}, \delta {\Phi}_0, \delta {\Phi}^{'}, \delta p_0\}$, such that
\begin{equation}
   \delta \Pi[\grave{\bm{\chi}}, \delta \grave{\bm{\chi}}]=\delta \Pi[\widetilde{\bm{\chi}}, \delta \widetilde{\bm{\chi}}]. 
   \label{eq:mod_var_format}
\end{equation}
Here, the magnetic potential in ${\mathcal{B}}_0^{'}$ is denoted as $\Phi^{'}$, and the Lagrange multiplier is expressed as follows:
\begin{equation}
p\left(\eta, \theta^\alpha\right)=p_0\left(\theta^\alpha\right)+\eta p_1\left(\theta^\alpha\right)+\mathcal{O}\left(\eta^2\right).
\end{equation}
The contribution of each integral in the first variation \eqref{eq:var_2} to the above modified format is discussed in detail in the following subsections. 
\subsection{Contribution to the first variation due to the total stress and the Maxwell stress}
\subsubsection{Integrals related to the total Piola stress tensor}
Taking into account the definition of  $\bm{\chi}_\mathrm{B}$, and following Equations ~\eqref{diff_vol} and \eqref{eq:undeformed_area_arb}, the first term in Equation~\eqref{eq:var_inside_body} which is the domain term related to the total Piola stress tensor can be written as
\begin{eqnarray}
-\int\limits_{{\mathcal{B}}_0}\mathrm{Div}\bm{P}\cdot\delta\bm{\chi}dV = -\int\limits_{S}\int\limits_{\eta}\mathrm{Div}\bm{P}\cdot\delta\bm{\chi}_\mathrm{B}d\eta dS 
= -\int\limits_{ S_\mathrm{m}}\int\limits_{\eta}\mathrm{Div}\bm{P}\cdot \delta\bm{\chi}_\mathrm{B}Md\eta d{\mathcal{S}}_\mathrm{m} \ ,
\end{eqnarray}
Here, $M$ is defined in Table \ref{demo-table2}, which outlines the surface and volume elements of the shell.
From Equations~\eqref{eq:div_total_stress} and \eqref{eq:shift_ref}, and noting that,
${\int\limits_{\eta} d\eta}=T \quad \text{and} \quad {\int\limits_{\eta} \eta d\eta}=0$,
\begin{eqnarray}
-\int\limits_{ S_\mathrm{m}}\int\limits_{\eta}\mathrm{Div}\bm{P}\cdot \delta\bm{\chi}_\mathrm{B}Md\eta d{\mathcal{S}}_\mathrm{m}
&=&\int\limits_{S_\mathrm{m}}\left[-T  {\bm{P}_0}_{, \alpha}{\bm{A}}^\alpha\cdot \delta\bm{r}-T\bm{P}_1 \bm{N}\cdot \delta\bm{r}+ \mathcal{O}\left(T^3\right)\right]dS_\mathrm{m} \ ,
  \label{eq:tot_stress_shell_var}
\end{eqnarray}
where
\begin{equation}
    \bm{P}_0=\bm{\mathcal{P}}_0-p_0\bm{F}^{-\mathrm{T}}_0,\quad \text{and} \quad
    \bm{P}_1=\bm{\mathcal{P}}_1-p_0\bm{F}^{-\mathrm{T}}_1-p_1\bm{F}^{-\mathrm{T}}_0 \ ,
    \label{eq:Lag_comp}
\end{equation}
with 
$\bm{\mathcal{P}}(\eta, \theta^\alpha)= \dfrac{\partial \Omega}{\partial\bm{F}}
  =\bm{\mathcal{P}}_0+\eta\bm{\mathcal{P}}_1+\bm{\mathcal{O}}(\eta^2)$.
The boundary term contribution related to the total Piola stress tensor at the top surface in Equation~\eqref{eq:var_inside_body} can be expressed as 
\begin{eqnarray}
\int\limits_{ S_\mathrm{t}}\bm{P} {\bm{N}}\cdot\delta\bm{\chi}dS_\mathrm{t}&=&
\int\limits_{ S_\mathrm{t}}\left[\bm{P}_0\ {\bm{N}}\cdot\delta \bm{r}+\frac{T}{2}\delta \lambda \bm{P}_0 \bm{N} \cdot \bm{n}+ \frac{T}{2} \lambda \bm{P}_0 \bm{N}\cdot \delta \bm{n}+\frac{T}{2}\bm{P}_1 \bm{N}\cdot\delta \bm{r}
+\mathcal{O}\left(T^2\right)\right]dS_\mathrm{t} \ .
\label{eq:top_total_sress}
\end{eqnarray}
Similarly,  for the bottom surface,
\begin{eqnarray}
-\int\limits_{ S_\mathrm{b}}\bm{P} {\bm{N}}\cdot\delta\bm{\chi}dS_\mathrm{b}
&=&\int\limits_{ S_\mathrm{b}}\left[-\bm{P}_0 \bm{N}\cdot\delta \bm{r}+\frac{T}{2}\delta \lambda \bm{P}_0 \bm{N} \cdot \bm{n}+ \frac{T}{2} \lambda \bm{P}_0 {\bm{N}}\cdot \delta \bm{n}+\frac{T}{2}\bm{P}_1 \bm{N}\cdot\delta \bm{r}
+\mathcal{O}\left(T^2\right)\right]dS_\mathrm{b} \ .
\label{eq:bottom_total_sress}
\end{eqnarray}
 In Equations ~\eqref{eq:top_total_sress} and \eqref{eq:bottom_total_sress}, the second and third integrals can be rewritten with the help of Equation~\eqref{eq:shiftor1} as
 \begin{subequations}
\begin{align}
 \int\limits_{ S_\mathrm{m}} \frac{T}{2}\delta \lambda \bm{P}_0 \bm{N} \cdot \bm{n} M\Big|_{\eta=\pm T/2}dS_\mathrm{m}&=\int\limits_{ S_\mathrm{m}}\left[\frac{T}{2}\delta \lambda \bm{P}_0 \bm{N} \cdot \bm{n}
  +\mathcal{O}\left(T^2\right)\right]
  dS_\mathrm{m} \ , \\
   \int\limits_{ S_\mathrm{m} } \frac{T}{2} \lambda \bm{P}_0 \bm{N} \cdot \delta \bm{n}
   M \Big|_{\eta=\pm T/2}dS_\mathrm{m}&=\int\limits_{ S_\mathrm{m}}\left[\frac{T}{2} \lambda \bm{P}_0 \bm{N}\cdot \delta \bm{n} 
  +\mathcal{O}\left(T^2\right)\right]
  dS_\mathrm{m} \ . 
\end{align}
\end{subequations}
Further, using Equations ~\eqref{eq:lateral_to_mid} and \eqref{eq:lateral_nor_to_mid}, the integral for the lateral boundary in Equation~\eqref{eq:var_inside_body} can be represented as follows:
\begin{equation}
\int\limits_{ S_\ell}\bm{P} \bm{\nu}_\mathrm{l}\cdot\delta\bm{\chi}dS_\ell
=\int\limits_{C_\mathrm{m}}\left[T\bm{P}_0 \bm{\nu} \cdot \delta \bm{r}+\mathcal{O}\left(T^3\right)\right]dl
=\int\limits_{C_\mathrm{m} \setminus C_\mathrm{m}^\mathrm{u} }\left[T \nu_{\alpha}\bm{P}_0 {\bm{A}}^{\alpha}   \cdot \delta \bm{r}+\mathcal{O}\left(T^3\right)\right]dl.
\end{equation}
\subsubsection{Integrals related to the Maxwell stress tensor}
Following eqn.~\eqref{eq:var_outside_body}, 
the terms concerning the Maxwell stress tensor at the inner boundaries of the free space can be written for the modified format as follows:
\begin{subequations}
\begin{align}
\int\limits_{ S_\mathrm{t}}{\bm{P}_\mathrm{M}}_{\mathrm{t}} {\bm{N}}\cdot\delta\bm{\chi}dS_\mathrm{t}
&=\int\limits_{ S_\mathrm{t}} {\bm{P}_\mathrm{M}}_\mathrm{t}
\bm{N}\cdot\delta \bm{r}dS_\mathrm{t} +
\int\limits_{ S_\mathrm{m}}\left[\frac{T}{2}\delta \lambda {\bm{P}_\mathrm{M}}_\mathrm{t}\bm{N} \cdot \bm{n}+ \frac{T}{2} \lambda {\bm{P}_\mathrm{M}}_\mathrm{t} \bm{N}\cdot \delta \bm{n}+\mathcal{O}\left(T^2\right)\right]dS_\mathrm{m} \ ,  \\
\int\limits_{ S_\mathrm{b}}{\bm{P}_\mathrm{M}}_{\mathrm{b}} {\bm{N}}\cdot\delta\bm{\chi}d{S}_\mathrm{b}
&= 
\int\limits_{ S_\mathrm{b}}{\bm{P}_\mathrm{M}}_\mathrm{b}\bm{N}\cdot\delta \bm{r}dS_\mathrm{b} -
\int\limits_{ S_\mathrm{m}}\left[\frac{T}{2}\delta \lambda {\bm{P}_\mathrm{M}}_\mathrm{b}\bm{N} \cdot \bm{n}+ \frac{T}{2} \lambda {\bm{P}_\mathrm{M}}_\mathrm{b} \bm{N}\cdot \delta \bm{n}+\mathcal{O}\left(T^2\right)\right]dS_\mathrm{m} \ ,  \\
\int\limits_{ S_\ell}{\bm{P}_\mathrm{M}}_\ell \cdot\delta\bm{\chi}dS_\ell
&=\int\limits_{C_\mathrm{m} \setminus C_\mathrm{m}^\mathrm{u} }\left[T{\nu}_{\alpha}{{\bm{P}}_{\mathrm{M}}}_0 {\bm{A}}^{\alpha}   \cdot \delta \bm{r}+\mathcal{O}\left(T^3\right)\right]dl \ . 
\end{align}
\end{subequations}
\subsubsection{Contribution arising from both stress tensors}
Now, considering the fictitious map $\bm{\chi}_\mathrm{F}$, the net contribution due to the Maxwell and total stress tensors to the modified variational form can be written as shown below:
\begin{eqnarray}
   &\textcolor{white}{=}&-\int\limits_{{\mathcal{B}}_0}\mathrm{Div}\bm{P}\cdot\delta\bm{\chi}dV
 + \int\limits_{ S_\mathrm{t}}\bm{P} {\bm{N}}\cdot\delta\bm{\chi}dS_\mathrm{t}
 -\int\limits_{ S_\mathrm{b}}\bm{P} {\bm{N}} \cdot\delta\bm{\chi}dS_\mathrm{b} 
 +\int\limits_{ S_\ell}\bm{P} \bm{\nu}_\ell \cdot\delta\bm{\chi}dS_\ell 
 -\int\limits_{{\mathcal{B}}_0^{'}}\mathrm{Div}\bm{P}_\mathrm{M}\cdot\delta\bm{\chi}dV\nonumber\\
 &\textcolor{white}{=}&
 -\int\limits_{ S_\mathrm{t}} {\bm{P}_\mathrm{M}}_\mathrm{t}
 {\bm{N}}\cdot\delta\bm{\chi}dS_\mathrm{t} 
 +\int\limits_{ S_\mathrm{b}} {\bm{P}_\mathrm{M}}_\mathrm{b}
 {\bm{N}}\cdot\delta\bm{\chi}dS_\mathrm{b} -\int\limits_{ S_\ell} {\bm{P}_\mathrm{M}}_\ell
 {\bm{\nu}}_\ell\cdot\delta\bm{\chi}dS_\ell \nonumber \\
    &=& \int\limits_{S_\mathrm{m}}\left[-T  {\bm{P}_0}_{, \alpha} {\bm{A}}^\alpha\cdot \delta\bm{r}
    -\frac{T}{2}\bm{P}_1 \bm{N}\cdot\delta \bm{r}
    +T\delta \lambda \grave{\bm{P}} \bm{N} \cdot \bm{n}+ T \lambda \grave{\bm{P}} \bm{N}\cdot \delta \bm{n} \right]
dS_\mathrm{m} 
\nonumber\\
&\textcolor{white}{=}&
+\int\limits_{ S_\mathrm{t}}\left[\bm{P}_0 \bm{N}\cdot\delta \bm{r}+ \frac{T}{2}\bm{P}_1 \bm{N}\cdot\delta \bm{r}
-{\bm{P}_\mathrm{M}}_\mathrm{t} \bm{N}\cdot\delta \bm{r}\right]dS_\mathrm{t}
+\int\limits_{ S_\mathrm{b}}\left[-\bm{P}_0 \bm{N}\cdot\delta \bm{r}+ T\bm{P}_1 \bm{N}\cdot\delta \bm{r}
+{\bm{P}_\mathrm{M}}_\mathrm{b} \bm{N} \cdot\delta \bm{r}\right]dS_\mathrm{b} 
\nonumber \\
&\textcolor{white}{=}&+\int\limits_{C_\mathrm{m} \setminus C_\mathrm{m}^\mathrm{u} }T {\nu}_{\alpha}\left[\bm{P}_0-{\bm{P}_\mathrm{M}}_0\right] {\bm{A}}^{\alpha}   \cdot \delta \bm{r}dl 
-\int\limits_{{\mathcal{B}}_0^{'}}\mathrm{Div}\bm{P}_\mathrm{M}\cdot\delta\bm{\chi}_\mathrm{F}dV
\ ,
\label{eq:cont_stress_tensor_1}
\end{eqnarray}
with $\grave{\bm{P}}=\bm{P}_0-\overline{\bm{P}}_\mathrm{M}$ and $\overline{\bm{P}}_\mathrm{M}=\dfrac{{\bm{P}_\mathrm{M}}_\mathrm{t}+{\bm{P}_\mathrm{M}}_\mathrm{b}
 }{2}$. The third and fourth term in the integral over the mid-surface of the shell can be  rewritten by using Equation~\eqref{eq:undeformed_area_mid} as 
\begin{eqnarray}
    \int\limits_{ S_\mathrm{m}} T\delta \lambda \grave{\bm{P}} \bm{N} \cdot \bm{n}dS_\mathrm{m}=\int\limits_{ P}T\delta \lambda \grave{\bm{P}} \bm{N} \cdot \bm{n}{A}^{1/2} dP \ , \quad \text{and} \quad
     \int\limits_{ S_\mathrm{m}}T \lambda \grave{\bm{P}} \bm{N}\cdot \delta \bm{n}dS_\mathrm{m}=\int\limits_{ P} T \lambda \grave{\bm{P}} \bm{N}\cdot \delta \bm{n}{A}^{1/2} dP \ .
     \label{eq:total_stress_param}
\end{eqnarray}
Further, using Equations ~\eqref{parametric_der_christoffel}, \eqref{eq:var_def_nor}, and \eqref{eq:var_lambda}, Equation~\eqref{eq:total_stress_param}$_1$ can be expressed as 
\begin{eqnarray}
\int\limits_{ P}T\delta \lambda \grave{\bm{P}} \bm{N} \cdot \bm{n}{A}^{1/2} dP&=&-\int\limits_{ P} T \lambda  \left[\grave{\bm{P}} \bm{N} \cdot \bm{n}\right] \bm{a}^{\alpha}\cdot\delta {\bm{a}}_{\alpha}{A}^{1/2} dP, \nonumber \\
&=&-\int\limits_{ P}\left[{C}^{\alpha}{{A}}^{1/2}\right]_{,\alpha}dP
+\int\limits_{ S_\mathrm{m}}\left[ T \lambda  \left[\grave{\bm{P}}  \bm{N} \cdot \bm{n}\right] \bm{a}^{\alpha}\right]_{, \alpha}\cdot\delta \bm{r} dS_\mathrm{m}
+\int\limits_{S_\mathrm{m}} C^\alpha {\Gamma}^\beta_{\beta \alpha} dS_\mathrm{m},
\label{eq:total_stress_param_simp_1}
\end{eqnarray}
with ${C}^{\alpha}=T \lambda  \left[\grave{\bm{P}}\bm{N} \cdot \bm{n}\right] \bm{a}^{\alpha}\cdot\delta \bm{r}$, and the Christoffel symbols of the second kind
$\bm{\Gamma}$ is defined in Table \ref{demo-table1}.
Similarly, Equation~\eqref{eq:total_stress_param}$_2$ can be written as
\begin{equation}
\int\limits_{ P}T \lambda \grave{\bm{P}} \bm{N}\cdot \delta \bm{n}{A}^{1/2} dP
=
-\int\limits_{ P}\left[{D}^{\alpha}{A}^{1/2}\right]_{,\alpha}dP
+\int\limits_{S_\mathrm{m}}\left[T\lambda \left[\grave{\bm{P}} \bm{N}\cdot\bm{a}^\alpha\right]\bm{n}\right]_{, \alpha} \cdot \delta \bm{r}dS_\mathrm{m} 
+\int\limits_{S_\mathrm{m}} D^\alpha {\Gamma}^\beta_{\beta \alpha} dS_\mathrm{m}\ ,
\label{eq:total_stress_param_simp_2}
\end{equation}
 with ${D}^{\alpha}=T\lambda \left[\grave{\bm{P}} \bm{N}\cdot\bm{a}^\alpha\right]\bm{n} \cdot \delta \bm{r}$.
Moreover, following Equation~\eqref{eq:dom_to_bound}, and excluding the parts of the boundary where displacements are specified, one obtains
\begin{equation}
-\int\limits_{ P}\left[{C}^{\alpha}{A}^{1/2}\right]_{,\alpha}dP=-\int\limits_{C_\mathrm{m} \setminus C_\mathrm{m}^\mathrm{u}}C^\alpha \nu_\alpha dl \ ,
\quad \text{and} \quad
-\int\limits_{ P}\left[{D}^{\alpha}{A}^{1/2}\right]_{,\alpha}dP=-\int\limits_{C_\mathrm{m} \setminus C_\mathrm{m}^\mathrm{u}}D^\alpha \nu_\alpha dl \ .
\label{eq:total_stress_param_simp_3}
\end{equation}
\subsection{Contribution to the first variation due to the magnetic induction vector}
\subsubsection{Integrals related to the magnetic induction vector inside the shell}
In Equation~\eqref{eq:var_inside_body}, using Equation~\eqref{eq:div_of_mag_ind}, the domain term involving the magnetic induction vector can be written as
\begin{eqnarray}
-\int\limits_{{\mathcal{B}}_0}\mathrm{Div}\bm{\bbm{B}}\delta \Phi dV
&=&\int\limits_{ S_\mathrm{m}}\left[-T {\bm{\bbm{B}}}_{0,\alpha} \cdot {\bm{A}}^\alpha  \delta {\Phi}_0-T {\bm{\bbm{B}}}_1 \cdot \bm{N}  \delta {\Phi}_0
+\mathcal{O}\left(T^3\right)\right]dS_\mathrm{m}.
\end{eqnarray}
From Equations ~\eqref{eq:potential_thin_shell_assumption} and \eqref{eq:kirchoff_love_equivalent},   the corresponding boundary term  at the top surface is given by
\begin{eqnarray}
\int\limits_{ S_\mathrm{t}} {\bm{\bbm{B}}}\cdot {\bm{N}}\delta {\Phi}dS_\mathrm{t}
&=& \int\limits_{ S_\mathrm{t}} \left[{\bm{\bbm{B}}}_0 \cdot {\bm{N}}+ \frac{T}{2} {\bm{\bbm{B}}}_1 \cdot {\bm{N}} +\mathcal{O}\left({T}^2\right)\right]\delta {\Phi}_0 dS_\mathrm{t} 
\nonumber\\
&\textcolor{white}{=}&+
\int\limits_{ S_\mathrm{t}} \frac{1}{2}\left[{\bm{\bbm{B}}}_0\cdot {\bm{N}}+ \frac{T}{2} {\bm{\bbm{B}}}_1 \cdot {\bm{N}} +\mathcal{O}\left({T}^2\right)\right]\delta {\Phi}_\mathrm{t}^{'}dS_\mathrm{t}
\nonumber \\
&\textcolor{white}{=}& -\int\limits_{ S_\mathrm{t}} \frac{1}{2}\left[{\bm{\bbm{B}}}_0\cdot {\bm{N}}+ \frac{T}{2} {\bm{\bbm{B}}}_1\cdot {\bm{N}} +\mathcal{O}\left({T}^2\right)\right] \delta{\Phi}_\mathrm{b}^{'}dS_\mathrm{t} \ . 
\label{eq:mag_ind_top_cont}
\end{eqnarray}
The continuity of the magnetic potential at the shell boundaries is enforced, allowing $\Phi_1$ to be expressed as 
$
     {\Phi}_1\left(\theta^\alpha\right)
     =\dfrac{{\Phi}_\mathrm{t}^{'}-{\Phi}_\mathrm{b}^{'}}{T},
$ with 
 ${\Phi}_\mathrm{t}^{'}$ and 
 ${\Phi}_\mathrm{b}^{'}
$ as the potential at the top and bottom boundaries, respectively.
The first integral can be rewritten as follows:
\begin{eqnarray}
 &\textcolor{white}{=}&\int\limits_{ S_\mathrm{t}} \left[{\bm{\bbm{B}}}_0\cdot {\bm{N}}+ \frac{T}{2} {\bm{\bbm{B}}}_1\cdot {\bm{N}} +\mathcal{O}\left({T}^2\right)\right]\delta {\Phi}_0 dS_\mathrm{t} \nonumber \\
 &=&\int\limits_{ S_\mathrm{m}}\left[{\bm{\bbm{B}}}_0\cdot \bm{N}\delta {\Phi}_0+\frac{T}{2} {\bm{\bbm{B}}}_1\cdot \bm{N}\delta {\Phi}_0-T H {\bm{\bbm{B}}}_0\cdot {\bm{N}}\delta {\Phi}_0+ \mathcal{O}\left(T^2\right)\right]dS_\mathrm{m} \ , 
 \label{eq:mag_ind_shell_top_1}
\end{eqnarray}
and from Equation \eqref{eq:ref_top_bot_to_mid}, noting that,
\begin{eqnarray}
dS_\mathrm{t}= \left[1-T H+{\frac{T}{4}}^2 K\right]{\left[1+T H+{\frac{T}{4}}^2 K\right]}^{-1}dS_\mathrm{b}
=\left[1-2TH+\mathcal{O}\left(T^2\right)\right]dS_\mathrm{b},
\end{eqnarray}
one can write for the third integral
\begin{eqnarray}
 &\textcolor{white}{=}&-\int\limits_{ S_\mathrm{t}}\frac{1}{2} \left[{\bm{\bbm{B}}}_0\cdot \bm{N}+ \frac{T}{2} {\bm{\bbm{B}}}_1\cdot {\bm{N}} +\mathcal{O}\left({T}^2\right)\right] \delta {\Phi}_\mathrm{b}^{'}
 dS_\mathrm{t}
 \nonumber \\
 &=&\int\limits_{ S_\mathrm{b}} \frac{1}{2}\left[-{\bm{\bbm{B}}}_0\cdot \bm{N}\delta {\Phi}_\mathrm{b}^{'}- \frac{T}{2} {\bm{\bbm{B}}}_1\cdot \bm{N}\delta {\Phi}_\mathrm{b}^{'} +2 T H {\bm{\bbm{B}}}_0\cdot \bm{N}\delta {\Phi}_\mathrm{b}^{'}+\mathcal{O}\left({T}^2\right)\right] dS_\mathrm{b} \ . 
 \label{eq:mag_ind_shell_top_2}
\end{eqnarray}

Incorporating Equations~\eqref{eq:mag_ind_shell_top_1} and \eqref{eq:mag_ind_shell_top_2}, and omitting the subscripts t and b for the exterior magnetic potential, one obtains
\begin{eqnarray}
 \int\limits_{ S_\mathrm{t}} {\bm{\bbm{B}}}\cdot {\bm{N}}\delta {\Phi}dS_\mathrm{t}&=&
 \int\limits_{ S_\mathrm{m}}\left[{\bm{\bbm{B}}}_0\cdot \bm{N}\delta {\Phi}_0+\frac{T}{2} {\bm{\bbm{B}}}_1\cdot \bm{N}\delta {\Phi}_0-T H {\bm{\bbm{B}}}_0\cdot \bm{N}\delta {\Phi}_0+ \mathcal{O}\left(T^2\right)\right]dS_\mathrm{m} 
 \nonumber \\
 &\textcolor{white}{=}& +\int\limits_{ S_\mathrm{t}} \frac{1}{2}\left[{\bm{\bbm{B}}}_0\cdot \bm{N}\delta {\Phi}^{'}+ \frac{T}{2} {\bm{\bbm{B}}}_1\cdot \bm{N}\delta {\Phi}^{'} +\mathcal{O}\left({T}^2\right)\right] dS_\mathrm{t} 
 \nonumber \\
 &\textcolor{white}{+}&+\int\limits_{ S_\mathrm{b}} \frac{1}{2}\left[-{\bm{\bbm{B}}}_0\cdot \bm{N}\delta {\Phi}^{'}- \frac{T}{2} {\bm{\bbm{B}}}_1 \cdot \bm{N}\delta {\Phi}^{'} +2 T H {\bm{\bbm{B}}}_0\cdot \bm{N}\delta {\Phi}^{'}+\mathcal{O}\left({T}^2\right)\right] dS_\mathrm{b} \ .
 \label{eq:mag_ind_top}
\end{eqnarray}
Similarly, for the bottom surface of the shell,
\begin{eqnarray}
-\int\limits_{ S_\mathrm{b}} {\bm{\bbm{B}}}\cdot {\bm{N}}\delta {\Phi}dS_\mathrm{b}
&=&\int\limits_{ S_\mathrm{m}}\left[-{\bm{\bbm{B}}}_0\cdot {\bm{N}}\delta {\Phi}_0+\frac{T}{2} {\bm{\bbm{B}}}_1\cdot {\bm{N}}\delta {\Phi}_0-TH {\bm{\bbm{B}}}_0\cdot {\bm{N}}\delta {\Phi}_0+\mathcal{O}\left(T^2\right)\right]dS_\mathrm{m} 
\nonumber \\
&\textcolor{white}{=}&+\int\limits_{ S_\mathrm{b}} \frac{1}{2}\left[-{\bm{\bbm{B}}}_0\cdot {\bm{N}}\delta {\Phi}^{'}+ \frac{T}{2} {\bm{\bbm{B}}}_1 \cdot \bm{N}\delta {\Phi}^{'} +\mathcal{O}\left({T}^2\right)\right] dS_\mathrm{b} 
\nonumber \\
&\textcolor{white}{=}&+ \int\limits_{S_\mathrm{t}} \frac{1}{2}\left[{\bm{\bbm{B}}}_0\cdot \bm{N}\delta {\Phi}^{'}- \frac{T}{2} {\bm{\bbm{B}}}_1\cdot \bm{N}\delta {\Phi}^{'} +2TH {\bm{\bbm{B}}}_0\cdot \bm{N}\delta {\Phi}^{'}+\mathcal{O}\left({T}^2\right)\right] dS_\mathrm{t} \ . 
 \label{eq:mag_ind_bottom}
\end{eqnarray}
Note that the effect of the external field on the response of the soft magnetoelastic thin shell,  leads to integrals over the top, bottom, and mid-surfaces during the derivation of the reduced-order model.
In Equation~\eqref{eq:var_inside_body}, the contribution corresponding to the lateral surface of the shell can be expressed as
\begin{eqnarray}
\int\limits_{S_\ell} {\bm{\bbm{B}}}\cdot \bm{\nu}_\ell\delta \Phi dS_\ell
&=&\int\limits_{C_\mathrm{m}}\left[T{\bm{\bbm{B}}}_0\cdot \bm{\nu}\delta {\Phi}_0 + \mathcal{O}\left(T^3\right)\right]dl \ .
\end{eqnarray}
\subsubsection{Contribution arising from the magnetic induction vector}
In Equation~\eqref{eq:var_outside_body}, for the exterior magnetic induction, the integral over the lateral surface can be written as follows:
\begin{eqnarray}
-\int\limits_{S_\ell} {\bm{\bbm{B}}}_\ell^{'}
\cdot \bm{\nu}_\ell \delta {\Phi}dS_\ell
&=&-\int\limits_{C_\mathrm{m}}\left[T {\bm{\bbm{B}}}_0^{'}\cdot {\bm{\nu}}\delta {\Phi}_0 + \mathcal{O}\left(T^3\right)\right]dl \ .
\label{eq:mag_field_out_lateral}
\end{eqnarray} 
The total contribution resulting from the magnetic induction vector in the shell and free space in the modified variational form can now be expressed as 
\begin{eqnarray}
 &\textcolor{white}{=}&-\int\limits_{{\mathcal{B}}_0^{'}}\mathrm{Div} {\bm{\bbm{B}}}^{'} \delta {\Phi}dV
 +\int\limits_{ \partial {\mathcal{V}}_0} {\bm{\bbm{B}}}^{'}\cdot {\bm{N}}^{'}\delta {\Phi}dS
    -\int\limits_{ S_\mathrm{t}} {\bm{\bbm{B}}}_\mathrm{t}^{'}
    \cdot {\bm{N}}\delta \Phi dS_\mathrm{t} 
    +\int\limits_{ S_\mathrm{b}} {\bm{\bbm{B}}}_\mathrm{b}^{'}
    \cdot {\bm{N}}\delta \Phi dS_\mathrm{b} 
  - \int\limits_{ S_\ell} {\bm{\bbm{B}}}_\ell^{'}
  \cdot {\bm{\nu}}_\ell \delta \Phi dS_\ell \nonumber \\
 &\textcolor{white}{=}& -\int\limits_{{\mathcal{B}}_0}\mathrm{Div}{\bm{\bbm{B}}}\delta \Phi dV
 + \int\limits_{ S_\mathrm{t}} {\bm{\bbm{B}}} \cdot {\bm{N}} \delta \Phi dS_\mathrm{t} - \int\limits_{ S_\mathrm{b}} {\bm{\bbm{B}}} \cdot {\bm{N}}\delta \Phi dS_\mathrm{b}+\int\limits_{ S_\ell} {\bm{\bbm{B}}} \cdot {\bm{\nu}}_\ell \delta \Phi dS_\ell
  \nonumber \\
  &=&-\int\limits_{{\mathcal{B}}_0^{'}}\mathrm{Div} {\bm{\bbm{B}}}^{'}\delta {\Phi}^{'} dV
  +\int\limits_{ \partial {\mathcal{V}}_0} {\bm{\bbm{B}}}^{'}\cdot {\bm{N}}^{'}\delta {\Phi}^{'}dS \nonumber\\
   &\textcolor{white}{=}&
   +\int\limits_{ S_\mathrm{t}} \left[{\bm{\bbm{B}}}_0 \cdot \bm{N}+TH {\bm{\bbm{B}}}_0 \cdot \bm{N}-{\bm{\bbm{B}}}_\mathrm{t}^{'}\cdot \bm{N}\right] \delta {\Phi}^{'}dS_\mathrm{t} 
    + \int\limits_{ S_\mathrm{b}} \left[-{\bm{\bbm{B}}}_0 \cdot \bm{N}+TH {\bm{\bbm{B}}}_0 \cdot \bm{N}+{\bm{\bbm{B}}}_\mathrm{b}^{'}\cdot \bm{N}\right] \delta {\Phi}^{'} dS_\mathrm{b}
   \nonumber \\
   &\textcolor{white}{=}&
    +\int\limits_{ S_\mathrm{m}}\left[-T {\bm{\bbm{B}}}_{0,\alpha} \cdot {\bm{A}}^\alpha  -2TH {\bm{\bbm{B}}}_0\cdot {\bm{N}}\right]\delta {\Phi}_0 dS_\mathrm{m}
   +\int\limits_{C_\mathrm{m}}\left[T {\bm{\bbm{B}}}_0 \cdot \bm{\nu}-T{\bm{\bbm{B}}}_0^{'}\cdot \bm{\nu}\right]\delta {\Phi}_0 dl \ .
   \label{eq:total_cont_mag_ind}
\end{eqnarray}
\subsection{Contribution to the first variation due to  external loads}
The terms related to the external mechanical loads, as they appear in Equation~\eqref{eq:var_2}, will now be expanded upon.
\subsubsection{Integrals related to externally applied pressure}
In the present work, a noteworthy aspect is the differentiation of the applied pressures at the top and bottom surfaces of the shell structure, instead of directly considering them on the mid-surface during the derivation of the shell system of equations. From Equations ~\eqref{eq:ref_top_bot_to_mid} and \eqref{eq:def_surface_top_bottom}, the virtual work due to the external pressure at the top  surface of the shell is given by 
\begin{eqnarray}
\int\limits_{ s_\mathrm{t}}\bm{p}_\mathrm{t}\cdot \delta\bm{\chi}ds_\mathrm{t}&=&-\int\limits_{ S_\mathrm{t}}p_\mathrm{t}\bm{n}\cdot\delta\bm{\chi}_\mathrm{B}\Big|_{\eta=T/2}\ 
\widehat{\mu}\Big|_{\eta=T/2}
   {\widehat{a}}^{1/2} 
   dS_\mathrm{t} \ ,
   \label{eq:press_top_vir}
\end{eqnarray}
where 
\begin{eqnarray}
   \widehat{\mu}=\dfrac{\mu}{M}= 
   \left[1-2\eta \lambda h+\eta^2 \lambda^2 \kappa\right]
   {\left[1-2\eta H+\eta^2 K\right]}^{-1}
   =\left[1-2\eta\left[\lambda h + H\right]+\mathcal{O}\left(\eta^2\right)\right].
\end{eqnarray}
Using the expression for $\widehat{\mu}$, and taking into account that $\bm{n}\cdot\bm{n}=1$, $\bm{n}\cdot\delta\bm{n}=0$, and $\lambda {\widehat{a}}^{1/2} =1$, Equation~\eqref{eq:press_top_vir} can be rewritten as
\begin{eqnarray}
\int\limits_{ s_\mathrm{t}}\bm{p}_\mathrm{t}\cdot \delta\bm{\chi}ds_\mathrm{t}
&=& \int\limits_{ S_\mathrm{t}}\left[-{\lambda}^{-1}p_\mathrm{t}\bm{n}\cdot\delta \bm{r}+\bm{a}^{\alpha}\cdot \delta \bm{a}_{\alpha} \frac{T}{2}p_\mathrm{t}+T \left[ h + \lambda^{-1} H\right] p_\mathrm{t}\bm{n}\cdot\delta \bm{r}+\mathcal{O}\left(T^2\right)\right] dS_\mathrm{t} \ .
\end{eqnarray}
Similarly, the virtual work due to the external pressure at the bottom surface of the thin shell can be expressed  as
\begin{eqnarray}
\int\limits_{ s_\mathrm{b}}\bm{p}_\mathrm{b}\cdot \delta\bm{\chi}ds_\mathrm{b}
   &=&\int\limits_{ S_\mathrm{b}}\left[{\lambda}^{-1}p_\mathrm{b}\bm{n}\cdot \delta \bm{r}+\bm{a}^{\alpha}\cdot \delta \bm{a}_{\alpha} \frac{T}{2}p_\mathrm{b}+T\left[ h + \lambda^{-1} H\right]p_\mathrm{b}\bm{n}\cdot \delta \bm{r}+\mathcal{O}\left(T^2\right)\right]dS_\mathrm{b} \ .
\end{eqnarray}
Therefore, using Equation~\eqref{eq:undeformed_area_arb},
\begin{eqnarray}
\int\limits_{ s_\mathrm{t}}\bm{p}_\mathrm{t}\cdot \delta\bm{\chi}ds_\mathrm{t}+\int\limits_{ s_\mathrm{b}}\bm{p}_\mathrm{b}\cdot \delta\bm{\chi}ds_\mathrm{b}&=&
 \int\limits_{ {\mathcal{S}}_\mathrm{t}}\left[-{\lambda}^{-1}p_t\bm{n}\cdot\delta \bm{r}
 +T \left[ h + \lambda^{-1}H\right] p_t\bm{n}\cdot\delta \bm{r}+\mathcal{O}\left(T^2\right)\right] dS_\mathrm{t}
 +
\nonumber\\
&\textcolor{white}{=}&\int\limits_{ S_\mathrm{b}}\left[{\lambda}^{-1}p_\mathrm{b}\bm{n}\cdot \delta \bm{r}
+T\left[ h + \lambda^{-1} H\right]p_\mathrm{b}\bm{n}\cdot \delta \bm{r}+\mathcal{O}\left(T^2\right)\right]dS_\mathrm{b}
+
\nonumber\\
&\textcolor{white}{=}&\int\limits_{ S_\mathrm{m}}\left[
\bm{a}^{\alpha}\cdot \delta \bm{a}_{\alpha} T\overline{p}
+\mathcal{O}\left(T^2\right)\right]dS_\mathrm{m} \ , 
\label{eq:cont_press_1}
\end{eqnarray}
with 
$\overline{p}=\dfrac{p_t+p_b}{2}$,
and the integral over the mid-surface excluding the higher-order terms can be further simplified as
\begin{eqnarray}
\int\limits_{ S_\mathrm{m}}\bm{a}_{\alpha}\cdot \delta \bm{a}^{\alpha} T\overline{p} dS_\mathrm{m}
&=&\int\limits_{C_\mathrm{m}\setminus C_\mathrm{m}^\mathrm{u}} T\overline{p} \bm{a}^{\alpha}  {\nu}_{\alpha} \cdot \delta \bm{r}dl
-\int\limits_{ S_\mathrm{m}}\left[T\overline{p} \bm{a}^{\alpha} \right]_{, \alpha}\cdot \delta \bm{r}dS_\mathrm{m} 
- \int\limits_{ S_\mathrm{m}} T\overline{p} \bm{a}^{\alpha} {\Gamma}^\beta_{\beta \alpha}\cdot\delta \bm{r}dS_\mathrm{m} \ .
\label{eq:pressure_grren_theom}
\end{eqnarray}
\subsubsection{Integral related to the dead load traction}
For the lateral surface of the shell, the contribution due to the dead load traction applied at the bounding curve of the mid-surface can be written as
\begin{eqnarray}
\int\limits_{ C_\mathrm{m} \setminus C_\mathrm{m}^\mathrm{u}}\bm{t}_\ell \cdot\delta\bm{\chi}dl =\int\limits_{ C_\mathrm{m} \setminus C_\mathrm{m}^\mathrm{u}}\bm{t}_\ell\cdot\delta\bm{\chi}_\mathrm{B}\Big|_{\eta=0}dl
=\int\limits_{ C_\mathrm{m} \setminus C_\mathrm{m}^\mathrm{u}}\bm{t}_\ell\cdot\delta \bm{r}dl \ .
\label{eq:dead_load_trac_cont}
\end{eqnarray}
\subsubsection{Integral related to the body force}
The virtual work due to the body force per unit volume can be further simplified as
\begin{eqnarray}
\int\limits_{{\mathcal{B}}_0}\bm{\mathfrak{B}}\cdot \delta\bm{\chi} dV
&=&\int\limits_{ S_\mathrm{m}}\left[T\bm{\mathfrak{B}}_0\cdot \delta \bm{r}+\mathcal{O}\left(T^3\right)\right]dS_\mathrm{m} \ ,
\label{eq:body_force_cont}
\end{eqnarray}
with $\bm{\mathfrak{B}} (\eta, \theta^\alpha)=\bm{\mathfrak{B}}_0+\eta \bm{\mathfrak{B}}_1+\bm{\mathcal{O}}(\eta^2)$ in ${\mathcal{B}}_0$.
\subsubsection{Net contribution due to the external loads}
The applied magnetic induction, as defined by Equation~\eqref{eq:mag_load}, along with the overall role of external mechanical loads on the modified variational framework, is considered. Consequently, the combined effect of the external stimulus can be expressed as 
\begin{eqnarray}
&\textcolor{white}{=}&-\int\limits_{ S_\mathrm{t}}\bm{p}_\mathrm{t}\cdot \delta\bm{\chi}dS_\mathrm{t}-\int\limits_{ S_\mathrm{b}}\bm{p}_\mathrm{b}\cdot \delta\bm{\chi}dS_\mathrm{b}-\int\limits_{{\mathcal{B}}_0}\bm{\mathfrak{B}}\cdot \delta\bm{\chi} dV-\int\limits_{C_\mathrm{m} \setminus C_\mathrm{m}^\mathrm{u}}\bm{t}_\ell \cdot\delta\bm{\chi}dl -\int\limits_{\partial {\mathcal{V}}_0} {\bm{\bbm{B}}}_\mathrm{e} \cdot {\bm{N}}^{'}\delta \Phi dS\nonumber \\
&=&\int\limits_{ S_\mathrm{t}}\left[{\lambda}^{-1}p_\mathrm{t}\bm{n}
 -T \left[ h + \lambda^{-1}H\right] p_\mathrm{t}\bm{n}\right]\cdot\delta \bm{r} dS_\mathrm{t}
-\int\limits_{ S_\mathrm{b}}\left[{\lambda}^{-1}p_\mathrm{b}\bm{n}
+T\left[ h + \lambda^{-1}H\right]p_\mathrm{b}\bm{n}\right]\cdot\delta \bm{r}dS_\mathrm{b}
\nonumber\\
&\textcolor{white}{=}&+\int\limits_{S_\mathrm{m}}\left[
\left[T\overline{p} \bm{a}^{\alpha} \right]_{, \alpha} +
T\overline{p} \bm{a}^{\alpha} {\Gamma}^\beta_{\beta \alpha}-T\bm{\mathfrak{B}}_0\right]\cdot \delta \bm{r}dS_\mathrm{m} 
-\int\limits_{C_\mathrm{m}\setminus C_\mathrm{m}^\mathrm{u}}\left[T\overline{p} {\nu}_{\alpha} \bm{a}^{\alpha}+\bm{t}_\ell \right]\cdot \delta \bm{r}dl \nonumber \\
&\textcolor{white}{=}&-\int\limits_{\partial {\mathcal{V}}_0} {\bm{\bbm{B}}}_\mathrm{e} \cdot {\bm{N}}^{'}\delta \Phi^{'} dS \ .
\label{eq:total_cont_ext}
\end{eqnarray}
\subsection{Contribution to the first variation due to the incompressibility constraint}
For the volume-preserving magnetoelastic body, when expanding the Lagrange multiplier along the thickness of the thin shell and considering only the first-order terms with respect to the through-thickness parameter in the modified variational form, the contribution arising from the incompressibility constraint can be expressed as follows:
\begin{eqnarray}
    \int\limits_{{\mathcal{B}}_0}\delta p \left[J-1\right]dV
    &=&\int\limits_{ S_\mathrm{m}}\left[T\left[J_0-1\right]+\mathcal{O}\left(T^3\right)\right]\delta p_0 dS_\mathrm{m} \ .
    \label{eq:incompressibility_cont}
\end{eqnarray}
\section{Governing equations for the Kirchhoff-Love magnetoelastic shell and accompanying free space}
The equations for a nonlinear magnetoelastostatic Kirchhoff-Love thin shell are now derived using the modified variational form. In Section  5, the contributions of the stress tensors, external loads, and magnetic induction vector to the modified variational format were determined. By adding Equations~\eqref{eq:cont_stress_tensor_1}, \eqref{eq:total_cont_mag_ind}, \eqref{eq:total_cont_ext}, and \eqref{eq:incompressibility_cont}, the variation of the total potential energy of the system is obtained. The state of magnetoelastic equilibrium is obtained by considering the variable $\delta \widetilde{\bm{\chi}}$ as arbitrary, which must correspond to an extremum of $\delta \Pi$. In other words, the first variation of the potential energy functional must be zero.

Now, by the arbitrary variation $\delta \bm{r}$, the shell-system of equations are obtained as follows:
\begin{subequations}
\begin{align}
     - T  {\bm{P}_0}_{, \alpha} {\bm{A}}^\alpha -T\bm{P}_1 \bm{N}
    +\left[T \lambda  \left[\grave{\bm{P}}\bm{N} \cdot \bm{n}\right] \bm{a}^{\alpha}\right]_{, \alpha}
     +
    \left[T\lambda \left[\grave{\bm{P}} \bm{N} \cdot \bm{a}^\alpha\right]\bm{n}\right]_{, \alpha} \nonumber \\
    + T \lambda  \left[\grave{\bm{P}} \bm{N} \cdot \bm{n}\right] \bm{a}^{\alpha} {\Gamma}^\beta_{\beta \alpha}
+T\lambda \left[\grave{\bm{P}} \bm{N} \cdot\bm{a}^\alpha\right]\bm{n} {\Gamma}^\beta_{\beta \alpha}
+\left[T\overline{p} \bm{a}^{\alpha} \right]_{, \alpha} +T\overline{p} \bm{a}^{\alpha} {\Gamma}^\beta_{\beta \alpha}
-T\bm{\mathfrak{B}}_0 &=\bm{0} \quad  \forall \mbf{X} \in S_\mathrm{m}, \label{eq:mech_dom} \\
T\left[{\bm{P}_0}-{\bm{P}_\mathrm{M}}_{0}\right] {\bm{A}}^{\alpha} {\nu}_{\alpha}
 -T \lambda  \left[\grave{\bm{P}} \bm{N} \cdot \bm{n}\right] \bm{a}^{\alpha} {\nu}_\alpha 
-T\lambda \left[\grave{\bm{P}} \bm{N}\cdot\bm{a}^\alpha\right]\bm{n} {\nu}_\alpha -T\overline{p}  \bm{a}^{\alpha} {\nu}_{\alpha}-\bm{t}_\ell&=\bm{0} \quad  \forall \mbf{X} \in C_\mathrm{m}\setminus C_\mathrm{m}^\mathrm{u} , \label{eq:mech_lat_bound} \\
\left[\bm{P}_0 -{\bm{P}_\mathrm{M}}_\mathrm{t}\right]\bm{N}+ \frac{T}{2}\bm{P}_1 {\bm{N}}
+
{\lambda}^{-1}p_t\bm{n}
 -T \left[ h + \lambda^{-1}H\right] p_\mathrm{t}\bm{n}&=\bm{0} \quad  \forall \mbf{X} \in S_\mathrm{t},
 \label{eq:mech_top} \\
  -\left[\bm{P}_0 -{\bm{P}_\mathrm{M}}_\mathrm{b} \right]\bm{N}+ \frac{T}{2}\bm{P}_1 \bm{N}
-
{\lambda}^{-1}p_\mathrm{b}\bm{n}
-T\left[ h + \lambda^{-1}H\right]p_\mathrm{b}\bm{n}&=\bm{0} \quad  \forall \mbf{X} \in S_\mathrm{b}.
\label{eq:mech_bot}
\end{align}
\end{subequations}

Also, considering the arbitrary variations $\delta \Phi_0$ and $\delta \Phi'$ in the shell and the free space, respectively, the equations obtained are given by
\begin{subequations}
\begin{align}
    -T {\bm{\bbm{B}}}_{0,\alpha} \cdot {\bm{A}}^\alpha  -2TH {\bm{\bbm{B}}}_0 \cdot \bm{N}&=0 \quad  \forall \mbf{X} \in  S_\mathrm{m},
    \label{eq:midsurface_balance} \\ 
    \left[{\bm{\bbm{B}}}_0 -{\bm{\bbm{B}}}_0^{'}\right]\cdot \bm{\nu}&=0 \quad  \forall \mbf{X} \in C_\mathrm{m},
    \label{eqb:mag_lat_bound} \\
    \left[{\bm{\bbm{B}}}_0 -{\bm{\bbm{B}}}_\mathrm{t}^{'}\right]\cdot \bm{N}+TH {\bm{\bbm{B}}}_0 \cdot \bm{N}
    &=0
    \quad \forall \mbf{X} \in S_\mathrm{t},
    \label{eq:top_surf} \\
    -\left[{{\bm{\bbm{B}}}_0 -\bm{\bbm{B}}}_\mathrm{b}^{'}\right] \cdot \bm{N}+TH {\bm{\bbm{B}}}_0 \cdot \bm{N}
    &=0
   \quad \forall \mbf{X} \in S_\mathrm{b}, 
   \label{eq:bottom_surf} 
   \end{align}
   \end{subequations}
   and
   \begin{subequations}
   \begin{align}
   \mathrm{Div}{\bm{\bbm{B}}}^{'}&=0\quad  \forall \mbf{X} \in {\mathcal{B}}_0^{'}, \\
   \big[ {\bm{\bbm{B}}}^{'} -{\bm{\bbm{B}}}_\mathrm{e} \big] \cdot {\bm{N}}^{'}&=0 \quad  \forall \mbf{X} \in \partial \mathcal{V}.
\end{align}
\end{subequations}
The arbitrary variation $\delta \bm{\chi}_\mathrm{F}$ and $\delta p_0$  leads to
\begin{equation}
    \mathrm{Div}\bm{P}_M=\bm{0} \quad \forall \mbf{X} \in {\mathcal{B}}_0^{'} \quad \quad \text{and} 
    \quad \quad J_0-1 =0 \quad \forall \mbf{X} \in S_\mathrm{m} .
    \label{eq:incomp_surf}
\end{equation}
Equation~\eqref{eq:incomp_surf}$_2$ returns  the incompressibility relation \eqref{eq:through_thickness} at the mid-surface, as derived in Section  3 .

Furthermore, by neglecting the higher-order terms, the condition on the magnetic potential at the surfaces of the shell structure, as given by Equation~\eqref{eq:ext_bound_mag_field}, can be rewritten as:
\begin{equation}
    -{\Phi_0}_{, \alpha} -\eta {\Phi_0}_{, \beta} B_\alpha^{\ \beta}-\eta {\left[\frac{\Phi_\mathrm{t}^{'}-\Phi_\mathrm{b}^{'}}{T}\right]}_{, \alpha}+\mathrm{Grad}\Phi^{'}\cdot \bm{A}_\alpha - \eta \mathrm{Grad}\Phi^{'} B_\alpha^{\ \beta} \cdot \bm{A}_\beta=0,
\end{equation}
with $\eta= \pm T/2$ at the top and bottom surfaces, respectively. 
\section{Finite inflation and magnetisation of a long cylindrical shell}
\begin{figure}
  \begin{center}
   \subfigure[]{
   \includegraphics[width=0.180\linewidth]{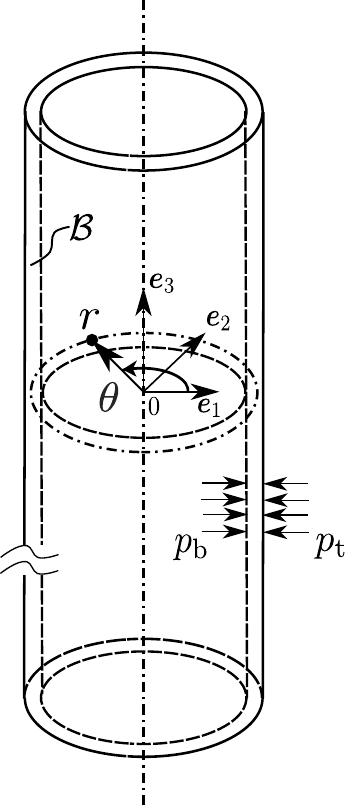}
   }  \quad \quad \quad \quad \quad \quad
   \subfigure[]{
   \includegraphics[width=0.264\linewidth]{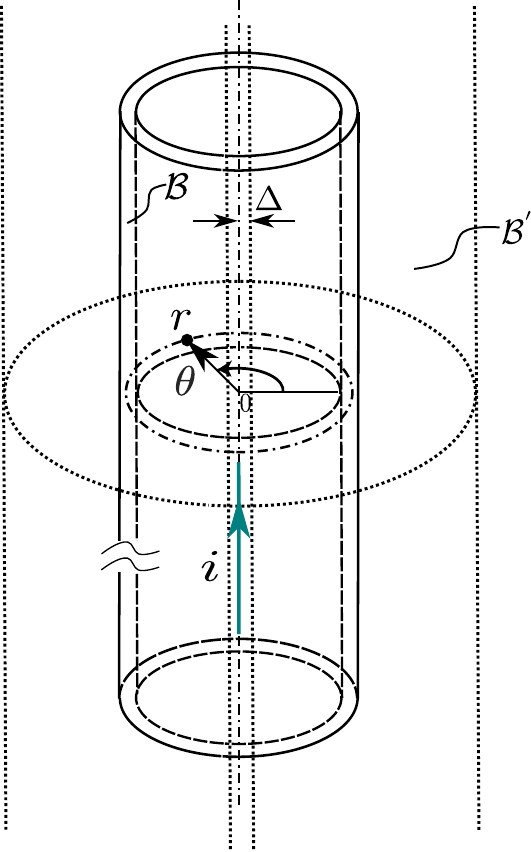}
   }
    \caption{Deformed configuration of an inflated infinite cylindrical shell depicting (a) Purely mechanical case with hydrostatic pressure applied at the inner and outer surfaces, and (b) Magnetoelastic case: A conductor carrying an electric current $i$ is placed along the axis of the cylinder, and the inner boundary of the free-space is at  an infinitesimal radial distance of $\Delta$, whereas the outer boundary of the free-space is at infinity. }
    \label{fig:fig5}
  \end{center}
\end{figure}

The main objective of this section is to illustrate via an example  how the equilibrium equations for a Kirchhoff-Love magnetoelastic thin shell, as introduced in Section  6, can be used to derive the response equations for the boundary-value problems at hand.
Consider the problem of an inflating infinite magnetoelastic thin cylindrical shell.
The body is subjected to two loading situations, as depicted in Figure \ref{fig:fig5}. The first scenario is  purely mechanical case where external pressures are applied at the inner and outer surfaces of the shell structure. The second scenario is the magnetoelastic case, with a wire carrying a current $i$ along the axis of the thin cylinder. The inner boundary of the free space is at an infinitesimal distance $\Delta$ from the wire while the outer boundary extends to infinity, as shown in Figure \ref{fig:fig5}.
The axisymmetric deformation of an infinite cylinder under a unit axial stretch is given by
\begin{subequations}
\begin{align}
\bm{r} &= \bm{R} + \bm{u} = \left[R+u\right]\bm{e}_{\rho} + Z\bm{e}_z = r\bm{e}_{\rho} + Z\bm{e}_z \\
\theta &= \Theta, \quad \quad
z = Z .
\end{align}
\end{subequations}
Here, $\theta$ and $z$ are the deformed coordinates corresponding to their azimuthal and axial counterparts in the reference configuration ($\Theta$ and $Z$, respectively). 
The unit vectors along the axial and radial directions are denoted by $\bm{e}_z$ and $\bm{e}_{\rho}$, respectively. 
Additionally, $R$ and $r$ represent the radius at the mid-surface of the cylindrical shell in the two configurations. 
The displacement vector is given by $\bm{u} = u(\rho)\bm{e}_{\rho}$. The covariant and contravariant vectors at the mid-surface in the two configurations, as well as the reference and deformed normals, are given by
\begin{eqnarray}
 &\textcolor{white}{=}&   {\bm{A}}_1=R\bm{e}_\theta, \quad {\bm{A}}_2=\bm{e}_z, \quad {\bm{A}}^1=\frac{1}{R}\bm{e}_\theta, \quad {\bm{A}}^2= \bm{e}_z, \nonumber \\
    &\textcolor{white}{=}&  {\bm{a}}_1=r\bm{e}_\theta, \quad {\bm{a}}_2=\bm{e}_z, \quad {\bm{a}}^1=\frac{1}{r}\bm{e}_\theta, \quad {\bm{a}}^2= \bm{e}_z,\nonumber \\
   &\textcolor{white}{=}&   \bm{n}=\bm{N}=\bm{e}_\rho, 
\end{eqnarray}
where $\bm{e}_\theta$ is the azimuthal unit vector.
The normal vectors in both the configurations coincide for the deforming cylinder, implying $\delta \bm{n}=\bm{0}$.
The components of the covariant and contravariant metric tensors at the mid-surface in the reference configuration are respectively
\begin{eqnarray}
    \left[{A}_{\alpha \beta}\right]=\begin{bmatrix}
{R}^2 & 0 \\
0 & 1 
\end{bmatrix} \quad  \text{and} \quad
 \left[{A}^{\alpha \beta}\right]=\begin{bmatrix}
{R}^{-2} & 0 \\
0 & 1 
\end{bmatrix},
\end{eqnarray}
and similarly, in the deformed configuration,
\begin{eqnarray}
    \left[{a}_{\alpha \beta}\right]=\begin{bmatrix}
{r}^2 & 0 \\
0 & 1 
\end{bmatrix} \quad  \text{and} \quad
 \left[{a}^{\alpha \beta}\right]=\begin{bmatrix}
{r}^{-2} & 0 \\
0 & 1 
\end{bmatrix},
\end{eqnarray}
along with the determinant of the covariant metric tensors at the mid-surface given by
\begin{equation}
    A={R}^2, \quad \text{and} \quad a={r}^2.
\end{equation}
From  Equation~\eqref{eq:incomp_surf}, one obtains
\begin{equation}
     J_0-1=0 \Rightarrow \lambda=\sqrt{\frac{A}{a}}=\lambda_\theta^{-1},
\end{equation}
with $\lambda_\theta=r/R$ as the azimuthal stretch. The  non-zero components of the curvature tensor at the mid-surface are 
\begin{equation}
    {B}_{1}^{\ 1}=-\frac{1}{R}, \quad \text{and} \quad {b}_{1}^{\ 1}=-\frac{1}{r}.
\end{equation}
Furthermore, 
\begin{equation}
    H=-\frac{1}{2 R}, \quad  h=-\frac{1}{2 r}, \quad \text{and} \quad  {\Gamma}^\beta_{\beta 1}={\Gamma}^\beta_{\beta 2}=0 .
\end{equation}
A generalised neo-Hookean  constitutive relation for magnetoelasticity \citep{dorfmann2014} is chosen where
\begin{equation}
    \Omega=\frac{\mu_\mathrm{s}}{4}
    \left[I_1-3\right]
    + \mu_0\left[\alpha I_4 + \beta I_5
    \right],
\end{equation}
and $ \beta=n\alpha$ and 
$\mu_\mathrm{s}$ is the shear modulus of the material.
The constants $\alpha$ and $\beta$ must be negative to ensure stability. Therefore,  for convenience, $\alpha=-1$, and $n \in \mathbb{R}^+$.
From Equation~\eqref{eq:cauchy_constitutive_general}, the total  Piola stress can be calculated as
\begin{equation}
    \bm{P}= \mu_\mathrm{s} \bm{F} + 2 \mu_0 \beta \bm{F}\bm{\bbm{H}}\otimes \bm{\bbm{H}}-p\bm{F^{-T}},
\end{equation}
and from Equation~\eqref{eq:mag_ind_rel_shell}, the magnetic field induction vector in the reference configuration of the shell is 
\begin{equation}
    \bm{\bbm{B}}=-2 \mu_0  \alpha \bm{\bbm{H}} -2 \mu_0  \beta \bm{C} \bm{\bbm{H}}.
\end{equation}
Now, the zeroth and first-order terms along the thickness of the shell of the total  Piola stress are given as 
\begin{subequations}
\begin{align}
    \bm{P}_0&=  \mu_{\mathrm{s}} \bm{F}_0 + 2 \mu_0 \beta \bm{F}_0 {\bm{\bbm{H}}}_0\otimes {\bm{\bbm{H}}}_0-p_0{\bm{F}_0}^{-T}, \\
    \bm{P}_1&=\mu_{\mathrm{s}} \bm{F}_1 +2\left[ \mu_0 \beta \bm{F}_0 {\bm{\bbm{H}}}_0\otimes {\bm{\bbm{H}}}_1+ \mu_0 \beta \bm{F}_0 {\bm{\bbm{H}}}_1\otimes {\bm{\bbm{H}}}_0+ \mu_0 \beta \bm{F}_1 {\bm{\bbm{H}}}_1\otimes {\bm{\bbm{H}}}_0\right]-\left[p_0 {\bm{F}_1}^{-T}-p_1 {\bm{F}_0}^{-T}\right],
\end{align}
\end{subequations}
and similarly,  the  components of the magnetic induction vector are
\begin{subequations}
\begin{align}
    {\bm{\bbm{B}}}_0&= -2 \mu_0 \alpha {\bm{\bbm{H}}}_0 -2 \mu_0 \beta \bm{C}_0 {\bm{\bbm{H}}}_0,  \\
{\bm{\bbm{B}}}_1&= -2 \mu_0 \alpha {\bm{\bbm{H}}}_1-2 \mu_0 \beta \bm{C}_0 {\bm{\bbm{H}}}_1
    -2\mu_0 \beta \bm{C}_1 {\bm{\bbm{H}}}_0 .
    \label{eq:mag_ind_cyilin}
\end{align}
\end{subequations}
Here  the applied magnetic field in the spatial configuration at a shell-point is 
$\bm{\bbm{h}}
=\dfrac{i}{2 \pi \left[r+\eta \lambda\right]}\bm{e}_\theta
=\bm{\bbm{h}}_0 + \eta \bm{\bbm{h}}_1=\dfrac{i \lambda}{2 \pi R}\bm{e}_\theta-\eta \dfrac{i \lambda^2}{2 \pi R^2}\bm{e}_\theta$ 
with $i \in {\mathbb{R}}^+$, and from the relation, $\bm{\bbm{H}}=\bm{F}^\mathrm{T}\bm{\bbm{h}}= { \bm{\bbm{H}}}_0+\eta  { \bm{\bbm{H}}}_1$, the following expressions are obtained:
\begin{equation}
  { \bm{\bbm{H}}}_0= \bm{F}_0^\mathrm{T} {\bm{\bbm{h}}}_0, \quad \text{and} \quad { \bm{\bbm{H}}}_1=
  \bm{F}_0^\mathrm{T} {\bm{\bbm{h}}}_1+\bm{F}_1^\mathrm{T} {\bm{\bbm{h}}}_0.
\end{equation}
The components of the deformation gradient and its inverse are calculated using Equations~\eqref{def_gradient} and \eqref{inv_def_gradient} as
\begin{subequations}
\begin{align}
 \bm{F}_0 &=\lambda \bm{e}_\rho \otimes \bm{e}_\rho + \lambda^{-1} \bm{e}_\theta \otimes \bm{e}_\theta 
    + \bm{e}_z\otimes \bm{e}_z,  \quad  \quad  \bm{F}_1=\frac{1}{R}\left[\lambda-\lambda^{-1}\right] \bm{e}_\rho \otimes \bm{e}_\rho, \\
    \bm{F}_0^{-\mathrm{T}} &=\lambda^{-1} \bm{e}_\rho \otimes \bm{e}_\rho + \lambda \bm{e}_\theta \otimes \bm{e}_\theta 
    + \bm{e}_z \otimes \bm{e}_z, \quad  \quad
     {\bm{F}_1}^{-\mathrm{T}}= \frac{1}{R}\left[\lambda-\lambda^{-3}\right] \bm{e}_\rho \otimes \bm{e}_\rho.
\end{align}
\end{subequations}
The expressions for the zeroth and first-order components of the deformation gradient and its inverse are required for evaluating 
the total Piola stress and the referential magnetic induction vector.
Therefore, considering the shell system of equations, specifically equations~\eqref{eq:mech_dom}, \eqref{eq:mech_top}, and \eqref{eq:mech_bot}, for the purely mechanical loading of the infinite soft cylinder in the absence of body force and pressure at the outer boundary, the response equation is given by
\begin{equation}
\frac{{p_\mathrm{b}}-{p_\mathrm{t}}}{\mu_\mathrm{s}} = \frac{T}{R}\left[\frac{p_b}{\mu_\mathrm{s}} + 1 - \lambda^4\right]\left[\frac{1}{2}\frac{T}{R}\left[1 + \lambda^2\right] + \frac{ \lambda^2}{2}\frac{T^2}{R^2} + 1\right]^{-1}.
\label{eq:press_diff}
\end{equation}
\begin{figure}
\centering
   \includegraphics[width=0.52\linewidth]{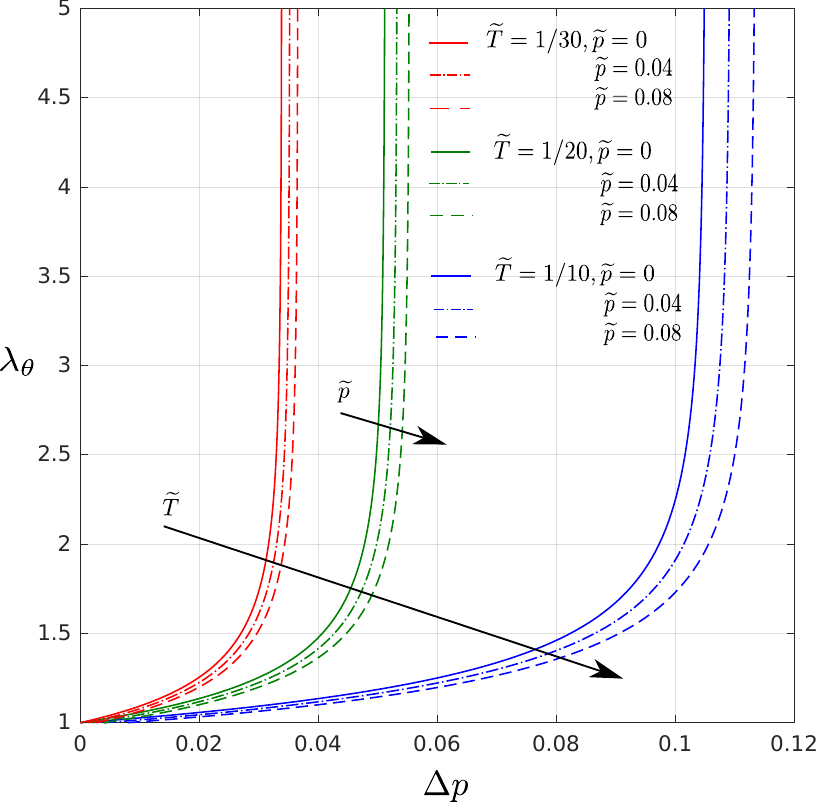}
    \caption{ Variation of the azimuthal stretch $\lambda_\theta$ with the dimensionless pressure difference between the bottom and top surfaces $\Delta p = [p_b - p_t]/\mu_s$ for shells of various thickness $\widetilde{T} = T/R$ and non-dimensional external pressure $\widetilde{p} = p_t/ \mu_s$.}
      \label{fig:purely_mech_1}
\end{figure}
\begin{figure}
    \centering
    \begin{tabular}{c c}
       \includegraphics[width=0.46\linewidth]{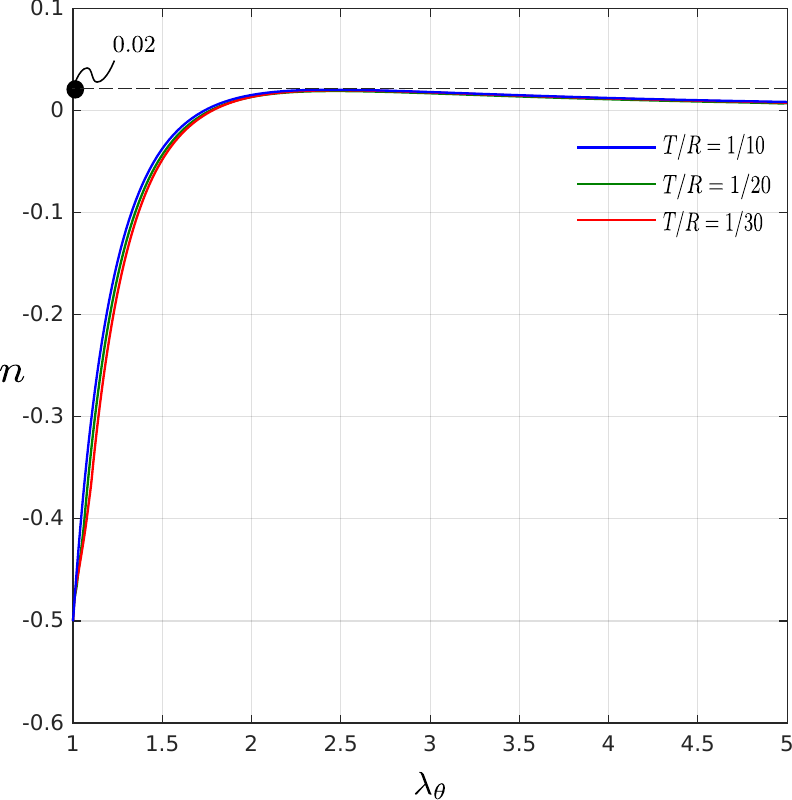}  &
       \includegraphics[width=0.45\linewidth]{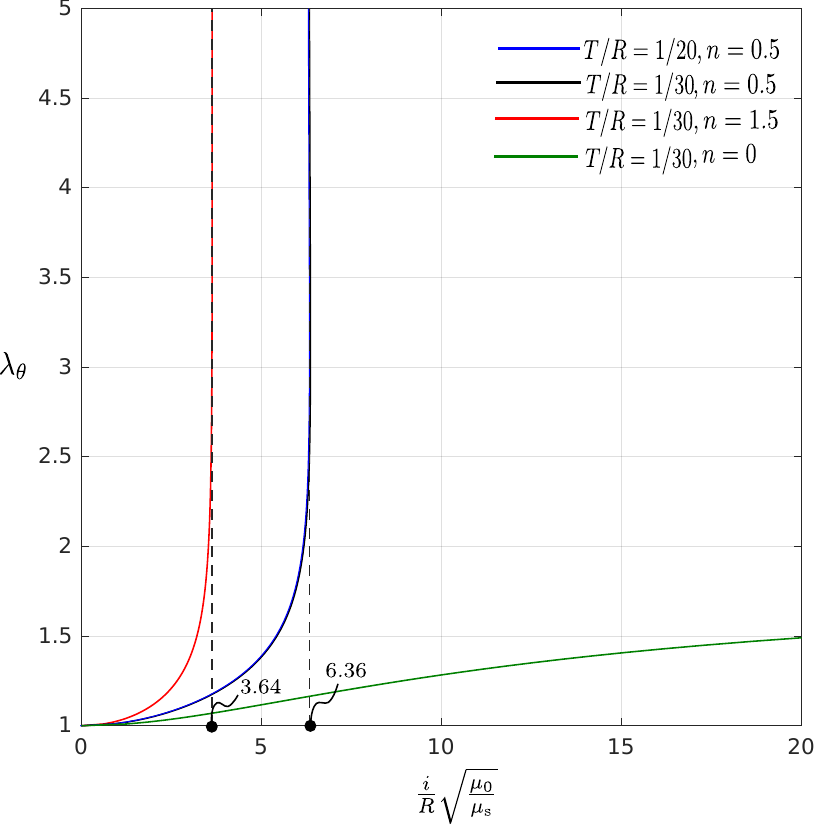} \\
       (a) & (b) \\[1ex]
       \includegraphics[width=0.485\linewidth]{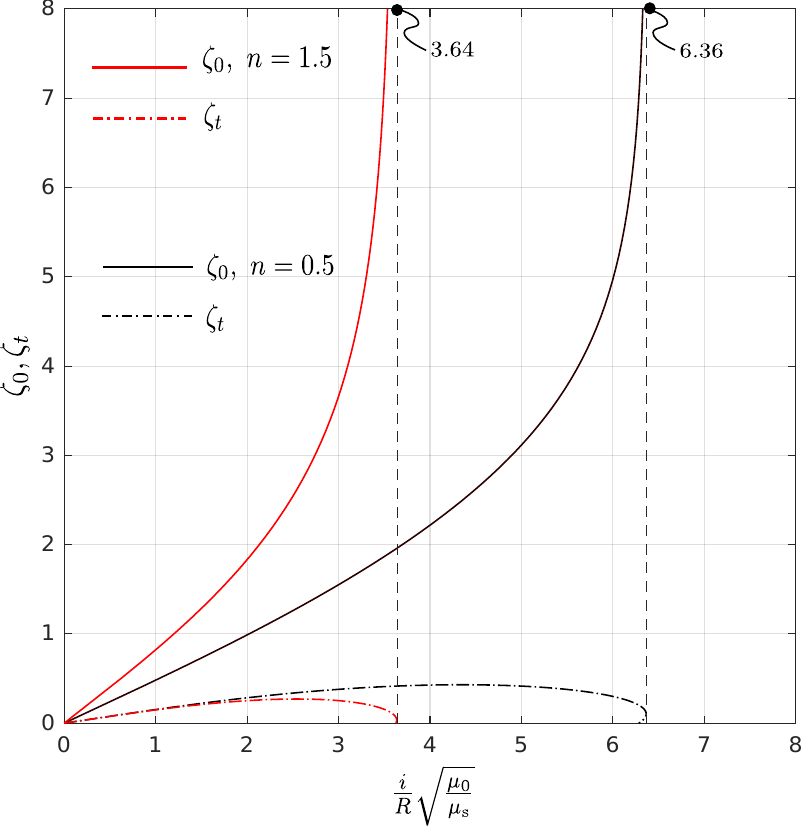}  &
       \includegraphics[width=0.46\linewidth]{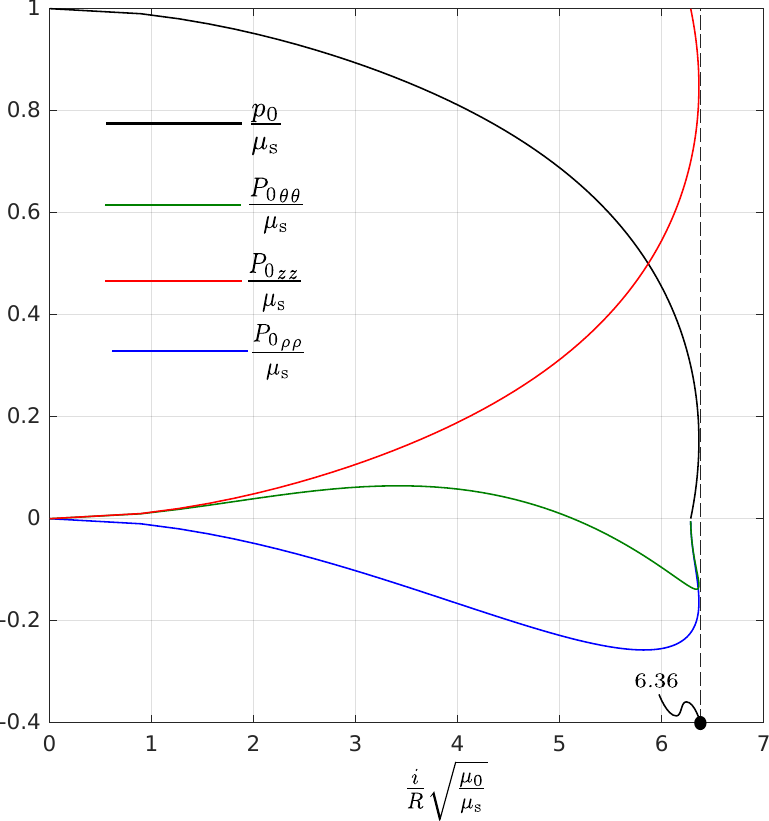} \\
       (c) & (d)
    \end{tabular}
    \caption{(a) Bounds on the constitutive parameter $n$ based on the inequality \eqref{eq:bound_n}. The curves for different thickness values ($T/R$) do not differ significantly from one another and it is found that choosing $n > 0.02$ ensures a physically consistent deformation for all $\lambda_\theta$ values. 
    (b) Variation of the inflation $\lambda_\theta$ of the cylindrical shell with the applied dimensionless magnetic loading.
    (c) Variation of the azimuthal magnetic induction at the shell mid, top, and bottom surfaces with the applied dimensionless magnetic loading.
    (d) Variation of the principal components of the zeroth-order  Piola stress and the Lagrange multiplier with the applied dimensionless magnetic loading for $T/R=1/30$ and $n=0.5$.
    }
    \label{fig: magnetic inflation graphs}
\end{figure}
This relationship is plotted in Figure \ref{fig:purely_mech_1} for different shell thickness values ($\widetilde{T}=T/R$) and external pressure values ($\widetilde{p}=p_t/\mu_s$).
As the pressure difference ($\Delta p = [p_b-p_t]/\mu_s$) between the inner and the external shell surface increases, the stretch $\lambda_\theta$ increases monotonically until a critical value of $\Delta p$ corresponding to a limit point instability is reached.
At this point massive changes in inflation occur for a minor change in the applied pressure.
Similar limit point instabilities have been  observed for inflation of thin hyperelastic shells as well as soft cylindrical cavities \citep{kiendl2015isogeometric, PhysRevLett.122.068003, MEHTA2022104031}.
The critical limit point pressure reduces as the shell thickness is reduced. 

We further demonstrate the distinction between  considering the pressure on top and bottom surfaces of the shell separately as opposed to the common convention of considering a pressure difference on the mid-surface.
The shell's response to applied pressure difference $\Delta p$ can significantly change by varying the pressure on the external surface $p_t/\mu_s$.
Reducing the shell thickness brings these response curves closer together, as observed from $T/R=1/30$ to $1/10$.

For the magnetoelastic deformation of the cylindrical shell due to an applied current along its axis, 
the equilibrium Equations~\eqref{eq:midsurface_balance}, \eqref{eq:top_surf}, and \eqref{eq:bottom_surf}, governing the magnetic induction vector are trivially satisfied. 
Furthermore, by considering the shell equilibrium Equations~\eqref{eq:mech_dom}, \eqref{eq:mech_top}, and \eqref{eq:mech_bot}, the following response relation for the system is obtained.
\begin{equation}
\frac{i }{R}\sqrt{\frac{\mu_0}{\mu_\mathrm{s}}}= 
 \sqrt{2} \pi\left[\lambda^4-1\right]^\frac{1}{2} \left[\frac{\left[1-\lambda^2\right] \left[4 \lambda^{-2}+\lambda^2 \dfrac{T^2}{R^2}\right] -8}{\left[4 \lambda^{-2}-\lambda^2 \dfrac{T^2}{R^2}\right]^2}
+\beta\right]^{-\frac{1}{2}}.
\label{eq:current_response}
\end{equation} 
Since, $\beta = -n$ and $\lambda<1$, it is evident that the condition,
\begin{equation}
 n > \frac{ \left[1-\lambda^2\right] \left[4 \lambda^{-2}+\lambda^2 \dfrac{T^2}{R^2} \right] -8}{\left[4 \lambda^{-2}-\lambda^2 \dfrac{T^2}{R^2}\right]^2},
    \label{eq:bound_n}
\end{equation}
must be satisfied by the constitutive parameter $n$ to ensure a physical deformation. 
This is further elaborated by plotting $n$ against the azimuthal stretch  for multiple $T/R$ values in Figure \ref{fig: magnetic inflation graphs}(a).
Based on this analysis, $n>0.02$ is necessary for an inflating cylindrical shell to ensure that the condition \eqref{eq:bound_n} is satisfied for all deformation states.

The deformation of the magnetoelastic cylinder based on Equation~\eqref{eq:current_response} is shown in 
Figure \ref{fig: magnetic inflation graphs}(b) 
for different $T/R$ and $n$ values. 
Application of magnetic field via the conductor causes the cylinder to inflate and the amount of inflation is higher for larger values of the coupling parameter $n$.
For a given $n\ne0$, there is a critical value of applied current $(i/R \sqrt{\mu_0/ \mu_s})$ beyond which the cylinder experiences rapid inflation, akin to a limit point instability \citep{Barham_2008, REDDY2017248}.
For $n=1.5$, this critical value is $i/R \sqrt{\mu_0/ \mu_s} \approx 3.64$, while for $n=0.5$, this is close to 6.36.
%
Notably, when $n=0$ for $T/R=1/30$, the cylinder exhibits slower inflation due to a weak magnetoelastic coupling, eventually saturating at $\lambda_\theta \approx 1.5$. This behaviour is alternatively explained from the requirement of the radial expansion to satisfy the imposed condition \eqref{eq:bound_n} on $n$, as presented in Figure \ref{fig: magnetic inflation graphs}(a). 
Furthermore, reducing the cylinder's thickness below a certain magnitude has a negligible effect for a given $n$, as indicated by the overlapping response curves for $T/R=1/20$ and $1/30$ at $n=0.5$.

The constitutive relation \eqref{eq:mag_ind_cyilin}$_1$ for the azimuthal component of the magnetic induction at the shell's mid-surface can be expressed as
\begin{equation}
     \zeta_0 :=\frac{{\bbm{B}_0}_\theta }{\sqrt{\mu_0 \mu_\mathrm{s}}}= -\pi^{-1}\left[\frac{i }{R}\sqrt{\frac{\mu_0}{\mu_\mathrm{s}}}\right]\left[\alpha+\beta \lambda^{-2}\right].
\end{equation}
The above indicates that $\zeta_0$ remains positive as the cylinder deforms.  
Additionally, the azimuthal components of the exterior magnetic induction at the outer and inner boundaries of the cylindrical shell, denoted as ${\bbm{B}_\mathrm{t}^{'}}_\theta$ and ${\bbm{B}_\mathrm{b}^{'}}_\theta$, respectively, are given by
\begin{subequations}
\begin{align}
 \zeta_t&:= \frac {{\bbm{B}_\mathrm{t}^{'}}_\theta }{\sqrt{\mu_0 \mu_\mathrm{s}}} =\pi^{-1}\left[\frac{i }{R}\sqrt{\frac{\mu_0}{\mu_\mathrm{s}}}\right]\left[\lambda + \frac{1}{2}\frac{T}{R}\left[\lambda-\lambda^3\right]\right]
   {\left[2\lambda^{-1}+\lambda \frac{T}{R}\right]}^{-1},  \\
   \zeta_b&:= \frac {{\bbm{B}_\mathrm{b}^{'}}_\theta }{\sqrt{\mu_0 \mu_\mathrm{s}}} =\pi^{-1}\left[\frac{i }{R}\sqrt{\frac{\mu_0}{\mu_\mathrm{s}}}\right]\left[\lambda - \frac{1}{2}\frac{T}{R}\left[\lambda-\lambda^3\right]\right]
  {\left[2\lambda^{-1}-\lambda \frac{T}{R}\right]}^{-1}.
\end{align}
\end{subequations}
The above expressions, along with the azimuthal magnetic induction at the mid-surface of the shell, are graphically presented in Figure \ref{fig: magnetic inflation graphs}(c) for $T/R=1/30$. 
Since $T/R = 1/30 \ll 1$ and $\lambda < 1$, $\zeta_t \approx \zeta_b$ and the two curves overlap, hence, only $\zeta_t$ is plotted.
The plot demonstrates that the magnetic induction at the mid-surface increases monotonically as the applied magnetic field (via the electric current) is increased.
Beyond a critical value of the applied current, the magnetic induction $\zeta_0$ increases abruptly similar to a limit point instability for the mechanical problem.
In the surrounding space, the magnetic induction at the shell boundaries exhibits an initial monotonic increase, followed by a gradual decrease, culminating in a sharp decline in magnitude. This decline occurs at relatively higher operational currents for smaller values of $n$, with slightly higher magnetic induction observed at this juncture for lower $n$. Additionally, for a specific $n$, the magnitude of the exterior magnetic induction at the inner surface is marginally greater than that at the outer boundary, although this difference diminishes as the cylinder inflates.

The radial, azimuthal, and axial components of the zeroth-order term of the total  Piola stress ($\bm{P}_0$) are
\begin{equation}
 \dfrac{{P_0}_{\rho \rho}}{\mu_\mathrm{s}}=\left[\lambda-\frac{p_0}{\mu_\mathrm{s}}\lambda^{-1}\right], \quad  \dfrac{{P_0}_{\theta \theta}}{\mu_\mathrm{s}}=\left[\lambda^{-1}+\frac{\lambda^{-1} \beta}{2 \pi^2 }{\left[\frac{i }{R}\sqrt{\frac{\mu_0}{\mu_\mathrm{s}}}\right]}^2-\frac{p_0}{\mu_\mathrm{s}}\lambda\right], \quad \text{and} \quad  \dfrac{{P_0}_{zz}}{\mu_\mathrm{s}}=\left[1-\frac{p_0}{\mu_\mathrm{s}}\right],
\end{equation}
respectively, where   
\begin{equation}
    \frac{p_0}{\mu_\mathrm{s}}=\left[\lambda^2-\frac{{P_\mathrm{Mt}}_{\rho \rho} + {P_\mathrm{Mb}}_{\rho \rho}}{2 \mu_\mathrm{s}} \lambda\right].
\end{equation}
The radial component of the Maxwell stress tensor in the surrounding space can be determined at the shell boundaries using Equation~\eqref{eq:Maxwell_stress_outside} and is given by
\begin{equation}
\dfrac{{P_\mathrm{Mt}}_{\rho \rho}}{\mu_\mathrm{s}}=-\dfrac{\lambda^{-1} \pi^{-2}}{2}{\left[\frac{i }{R}\sqrt{\frac{\mu_0}{\mu_\mathrm{s}}}\right]}^2 {\left[2 \lambda^{-1}+\lambda \frac{T}{R}\right]}^{-2} \quad \text{and} \quad \dfrac{{P_\mathrm{Mb}}_{\rho \rho}}{\mu_\mathrm{s}}=-\dfrac{\lambda^{-1} \pi^{-2}}{2} {\left[\frac{i }{R}\sqrt{\frac{\mu_0}{\mu_\mathrm{s}}}\right]}^2 {\left[2 \lambda^{-1}-\lambda \frac{T}{R}\right]}^{-2},
\end{equation}
at the outer and inner surfaces of the cylinder, respectively.
Figure \ref{fig: magnetic inflation graphs}(d) presents the principal components of $\bm{P}_0$ for $n=0.5$ and $T/R=1/30$.
The axial stress shows a monotonic increase with the applied magnetic loading until the limit point at $i/R \sqrt{\mu_0/ \mu_s} = 6.36$.
The radial component of the total Piola stress remains negative suggesting compression.
It drops to a minimum value of $-0.26$ before increasing slightly until the limit point.
The azimuthal component displays a non-monotonic behaviour.
It initially increases and remains positive for low magnetic loads but then starts to decrease and becomes negative for $i/R \sqrt{\mu_0/ \mu_s} > 5.2$.
It is notable that the radial stress is non-negligible compared to the other components, indicating the presence of inaccuracies when employing plane stress assumptions in modelling soft magnetoelastic shells.
Also, the zeroth-order component of the Lagrange multiplier is plotted, and it  decreases as the magnetoelastic cylindrical shell inflates.
The presence of compressive stresses in the azimuthal direction indicates a possibility of wrinkling instability, the analysis of which will form part of a future study.
\section{Concluding Remarks}
In this work, 
The governing equations for the large deformation of Kirchhoff-Love magnetoelastic thin shells have been rigorously derived. 
The free space in which the magnetostatic energy is bounded to finite volumes is accounted for.
The equilibrium equations have been obtained using the derived theory approach.
This point of departure was a
variational form for a three-dimensional continuum magnetoelastic body involving mechanical deformation, magnetic field, and a Lagrange multiplier in the presence of body force, dead-load traction along the bounding curve of the mid-surface, external pressures at the top and bottom surfaces, and an external magnetic field. 
Treating the shell as a stack of surfaces, the general deformation map in the body has been restated in terms of a point on the deformed mid-surface. This requires
an additional term that incorporates the through-thickness stretch and the deformed normal (i.e., the first director). By defining a new set of generalised solution variables and thereby modifying the variational form, the shell equilibrium equations have been obtained. The thickness variable has been separated from the surface parameters, and the field variables  expanded along the thickness of the thin shell. Additionally, the governing equations for the corresponding three-dimensional free space are derived.

The new formulation relies on a Kirchhoff-Love type kinematic assumption for the magnetic scalar potential thereby ensuring a consistent derivation of the governing equations.
The top and bottom surfaces of the shell are considered in addition to the mid-surface to consistently account for the magnetic field in vacuum.
This leads to a departure from the commonly used plane-stress assumption in thin shell theory.
Furthermore, a distinction between the hydrostatic pressure applied on the top and bottom surfaces has been considered.
Variations of the through-thickness stretch and the deformed normal introduces richness into the formulation together with additional complexity
through the distribution of the parametric derivative of the mid-surface position vector to the lateral bounding curve.
The novel magnetoelastic shell theory and implications of the factors 
have been
illustrated by analysing the inflation of a cylindrical magnetoelastic shell. 
Capabilities of the present theory to model large deformation and limit point instabilities have been demosntrated.
The possibility of wrinkling instabilities 
due to the presence of compressive in-plane stresses in the shell have been detailed.

The present analysis provides a new
perspective into a strongly-coupled 
shell system of equations, which is challenging to obtain due to strong kinematic and constitutive nonlinearities.
The geometrically exact formulation ensures a high level of accuracy.
The focus here is on formulating and demonstrating the capabilities of the derived equations.
The derivation from a variational formulation ensures that the theory is amenable for numerical implementation via the finite element method.
Details of the numerical implementation will be presented in a future contribution.

\section*{Acknowledgements}
This work was supported by the UK Engineering and Physical Sciences Research Council (EPSRC) grants EP/V030833/1 and EP/R008531/1, and a Royal Society grant IES/R1/201122.

\begin{appendices}
\renewcommand{\theequation}{A.\arabic{equation}}
\section{Geometry of a Kirchhoff--Love thin shell}
\subsection{The natural basis at the mid-surface}
The covariant basis vectors for the mid-surface in the reference and deformed configurations, respectively, can be expressed as
\begin{equation}
\bm{A}_\alpha = \frac{\partial \bm{R}}{\partial \theta^\alpha}, 
\quad \mathrm{and} \quad
\bm{a}_\alpha = \frac{\partial \bm{r}}{\partial \theta^\alpha}.
\end{equation}
Thus, the unit normal vectors in the two configurations are defined by
\begin{equation}
\bm{N} = \frac{{\bm{A}}_1 \times {\bm{A}}_2}{{A}^{1/2}}, \quad \text{and} \quad  \bm{n} = \frac{\bm{a}_1 \times \bm{a}_2}{{a^{1/2}}},
\label{eq:normal}
\end{equation}
where $A$ and $a$ are 
\begin{equation}
A=\norm{ {\bm{A}}_1 \times {\bm{A}}_2}^{2},\quad \text{and} \quad a = \norm{\bm{a}_1 \times \bm{a}_2}^{2}.
\end{equation}
Further, it can be shown that
\begin{equation}
  A= \mathrm{det}\left[{A}_{\alpha \beta}\right],\quad \text{and} \quad a = \mathrm{det}[{a}_{\alpha \beta}]. 
\end{equation}rm.
The covariant components of the metric tensor for the mid-surface points $\bm{R}$ and $\bm{r}$ are respectively given by
\begin{equation}
{A}_{\alpha \beta} ={\bm{A}}_\alpha \cdot {\bm{A}}_\beta, \quad \text{and} \quad {a}_{\alpha \beta} = \bm{{a}}_{\alpha } \cdot \bm{{a}}_{ \beta}.
\end{equation}
Also, the contravariant metric tensor components for the mid-surface are 
\begin{equation}
{A}^{\alpha \gamma} {A}_{\gamma \beta} = \delta^{\alpha}_{\beta}, \quad \text{and} \quad {a}^{\alpha \gamma} {a}_{\gamma \beta} = \delta^{\alpha}_{\beta},
\label{eq:co_and_contra_metric}
\end{equation}
where $\delta^{\alpha}_{\beta}$ denotes the Kronecker delta. Again,  ${\bm{A}}^{\alpha}$ and $\bm{a}^{\alpha}$ denote the contravariant basis vectors for the mid-surface in the two configurations, defined by 
\begin{equation}
{\bm{A}}^{\alpha} \cdot {\bm{A}}
_{\beta} = \delta^\alpha_\beta, \quad \text{and} \quad \bm{a}^{\alpha} \cdot \bm{a}_{\beta} = \delta^\alpha_\beta .
\label{eq:deform_tensor}
\end{equation} 
\subsection{The unit alternator and permutation symbol}
In general, for a surface tensor ${\bm{Q}}=Q_{\alpha \beta}{\bm{A}}^{\alpha} \otimes {\bm{A}}^{\beta}$, the surface inverse ${\bm{Q}}^{-1}$ defined from
\begin{equation}
    {\bm{Q}}^{-1}{\bm{Q}}={\bm{I}},
\end{equation}
with $\bm{I}={\bm{A}}^{\beta} \otimes {\bm{A}}_{\beta} = {\bm{A}}_{\beta} \otimes {\bm{A}}^{\beta} $ ($\bm{I}$ denotes  the projection onto the tangent plane of ${S}_{\mathrm{m}}$) has the contravariant components as
\begin{equation}
  {Q}_{\mathrm{inv}}^{\alpha \beta}=\frac{1}{Q} {e}^{\alpha \gamma} {Q}_{ \delta \gamma}{e}^{\beta \delta},
  \label{surf_inv}
\end{equation}
where $Q=\mathrm{det}\left[{Q}_{ \alpha \beta}\right]$, and  the so-called unit alternator given as  
\begin{equation}
    \left[e^{\alpha \gamma}\right]=\begin{bmatrix}
0 & 1 \\
-1 & 0 
\end{bmatrix}.
\end{equation}
Further, the permutation tensor is defined by
\begin{equation}
    \bm{E}={E}^{\alpha \beta} \ {\bm{A}}_{\alpha} \otimes {\bm{A}}_{\beta}= \frac{1}{{A}^{1/2}}{e}^{\alpha \beta} \ {\bm{A}}_{\alpha} \otimes {\bm{A}}_{\beta}, \quad \text{and} \quad  \bm{\varepsilon}={\varepsilon}^{\alpha \beta} \ \bm{a}_{\alpha} \otimes \bm{a}_{\beta}= \frac{1}{{a}^{1/2}}e^{\alpha \beta} \ \bm{a}_{\alpha} \otimes \bm{a}_{\beta},
    \label{eq:permutation_symbol_1}
\end{equation}
in the two configurations. In particular, Equation \eqref{surf_inv} yields
\begin{equation}
    {A}^{\alpha \beta}=\frac{1}{{A}} {e}^{\alpha \gamma} {A}_{\gamma \delta} {e}^{\beta \delta} \quad \text{and} \quad 
    a^{\alpha \beta}=\frac{1}{a} e^{\alpha \gamma}a_{\gamma \delta}e^{\beta \delta}.
    \label{alternate_exp_metric_1}
\end{equation}
Using the relation, $e^{\gamma \alpha }e_{\gamma \beta}=\delta^\alpha_\beta$, 
\begin{equation}
  {A}^{\alpha \beta} {e}_{\beta \gamma} {A}= {e}^{\alpha \beta}{A}_{ \beta \gamma}\quad \text{and} \quad
   {a}^{\alpha \beta} {e}_{\beta \gamma}a= {e}^{\alpha \beta} {a}_{ \beta \gamma},
\end{equation}
which can be further used to rewrite the permutation tensors as
\begin{equation}
    \bm{E}={E}_{\alpha \beta}\ {\bm{A}}^{\alpha} \otimes {\bm{A}}^{\beta}= {A}^{1/2} {e}_{\alpha \beta} \ {\bm{A}}^{\alpha} \otimes {\bm{A}}^{\beta}, \quad 
    \text{and} 
    \quad  \bm{\varepsilon}={\varepsilon}_{\alpha \beta}\ \bm{a}^{\alpha} \otimes \bm{a}^{\beta}= {a}^{1/2}e_{\alpha \beta} \ \bm{a}^{\alpha} \otimes \bm{a}^{\beta}.
    \label{eq:permutation_symbol_2}
\end{equation}
Again, multiplying Equations~\eqref{alternate_exp_metric_1}$_1$ and \eqref{alternate_exp_metric_1}$_2$ by ${A}_{\alpha \beta}$ and $a_{\alpha \beta}$, respectively, one obtains
\begin{equation}
   A=\frac{1}{2} {e}^{\alpha \gamma} {e}^{\beta \delta} {A}_{\alpha \beta} {A}_{\gamma \delta}, 
   \quad \text{and} \quad 
   a=\frac{1}{2} e^{\alpha \gamma} e^{\beta \delta} a_{\alpha \beta} a_{\gamma \delta}.
    \label{alternate_exp_det}  
\end{equation}
From the above, one can write
\begin{equation}
  {A}_{,\zeta}=  {e}^{\alpha \gamma} {e}^{\beta \delta} {A}_{\alpha \beta, \zeta} {A}_{\gamma \delta}, \quad \text{and} \quad 
  a_{,\zeta}= e^{\alpha \gamma} e^{\beta \delta} a_{\alpha \beta, \zeta} a_{\gamma \delta},
\end{equation}
which can be further rewritten as
\begin{equation}
      {A}_{,\zeta}= {A} {A}^{\alpha \beta}  {A}_{\alpha \beta, \zeta}, \quad \text{and} \quad 
  a_{,\zeta}= a a^{\alpha \beta}  a_{\alpha \beta, \zeta} ,
  \label{det_metric_parametric_der}
\end{equation}
by using Equations~\eqref{alternate_exp_metric_1}$_1$ and \eqref{alternate_exp_metric_1}$_2$. Now,
\begin{equation}
 {A}_{\alpha \beta, \zeta}={\bm{A}}
_{\alpha, \zeta}\cdot {\bm{A}}
_{\beta} + {\bm{A}}
_{\alpha}\cdot {\bm{A}}
_{\beta, \zeta} , \quad\text{and} \quad 
 {a}_{\alpha \beta, \zeta}={\bm{a}}
_{\alpha, \zeta}\cdot{\bm{a}}
_{\beta} + {\bm{a}}
_{\alpha}\cdot{\bm{a}}
_{\beta, \zeta}.
\end{equation}
Therefore, 
\begin{equation}
   {A}_{,\zeta}= 2{A} {\Gamma}^\alpha_{\alpha \zeta}, \quad \text{and} \quad {a}_{,\zeta}= 2{a} {\gamma}^\alpha_{\alpha \zeta},
   \label{parametric_der_christoffel}
\end{equation}
with the  Christoffel symbols of the second kind in the two configurations defined by 
\begin{equation}
  {\Gamma}^\alpha_{\zeta \gamma}= {\bm{A}}^{\alpha} \cdot {\bm{A}}_{\zeta, \gamma}, \quad \text{and} \quad  {\gamma}^\alpha_{\zeta \gamma}= {\bm{a}}^{\alpha} \cdot {\bm{a}}_{\zeta, \gamma}.
\end{equation}
\subsection{The natural basis at a shell-point}
A point $\bm{x}_{\mathrm{B}} \in \mathcal{B}$ can be written as 
\begin{equation}
\bm{x}_{\mathrm{B}}=\bm{r}+\eta \bm{d}, 
\end{equation} 
where $\bm{d}=\lambda \bm{n}$ and $\lambda = \dfrac{t}{T}$. 
The covariant basis vectors at a point  $\bm{X}_{\mathrm{B}}$  in the shell are given by 
\begin{eqnarray}
{\bm{G}}_{\alpha}=\frac{\partial \bm{X}_{\mathrm{B}}}{\partial \theta^{\alpha}}
=\frac{\partial \bm{R}}{\partial \theta^{\alpha}} + \eta \frac{\partial \bm{N}}{\partial \bm{R}} \frac{\partial \bm{R}}{\partial \theta^{\alpha}}
= \bm{M}{\bm{A}}_{\alpha},
\end{eqnarray}
where
\begin{equation}
    \bm{M}=\bm{I}-\eta \bm{K},
    \label{eq:shiftor1}
\end{equation}
with  
\begin{equation}
  \bm{K} =-\dfrac{\partial \bm{N}}{\partial \bm{R}}=-{\bm{N}}_{, \beta} \otimes {\bm{a}}^{\beta}.
\end{equation}
Again, the covariant basis vectors at a point  $\bm{x}_{\mathrm{B}}$  in the shell are 
\begin{eqnarray}
\bm{g}_{\alpha}=\frac{\partial \bm{x}_\mathrm{B}}{\partial \theta^{\alpha}}
=\frac{\partial \bm{r}}{\partial \theta^{\alpha}} +\eta \bm{n} \frac{\partial \lambda}{\partial \theta^{\alpha}}+\eta \lambda \frac{\partial \bm{n}}{\partial \bm{r}} \frac{\partial \bm{r}}{\partial \theta^{\alpha}}
= \bm{\mu}\bm{a}_{\alpha}.
\end{eqnarray}
Here  the long-wave assumption has been considered \citep{ kiendl2015isogeometric, liu2022computational}. That is,
\begin{equation}
  \lambda_{,\alpha} \approx 0 .
  \label{eq:long_wave}
\end{equation}
While this assumption is strong, it is reasonable since the thickness of the shell is typically very small, resulting in negligible out-of-plane shearing and localised necking.
Therefore, 
\begin{equation}
    \bm{\mu}=\bm{i}-\eta \lambda \bm{\kappa},
    \label{eq:shiftor2}
\end{equation}
where  $\bm{i}=  \bm{a}^{\beta} \otimes \bm{a}_{\beta}=\bm{a}_{\beta} \otimes \bm{a}^{\beta}$ denotes  the projection onto the tangent plane of $s_\mathrm{m}$, the  deformed counterpart of ${S}_\mathrm{m}$. Also,
\begin{equation}
    \bm{\kappa}=-\dfrac{\partial \bm{n}}{\partial \bm{r}}
    =-{\bm{n}}_{, \beta} \otimes {\bm{a}}^{\beta}.
\end{equation}
Again, ${\bm{N}}_{, \beta}$ and ${\bm{n}}_{, \beta}$ is given by
\begin{equation}
    {\bm{N}}_{, \beta}=-{B}_\beta^{ \ \gamma} \ {\bm{A}}_{\gamma}, \quad \text{and} \quad {\bm{n}}_{, \beta}=-{{b}}_\beta^{ \ \gamma} \ {\bm{a}}_{\gamma},
\end{equation}
with the surface curvature tensors in the two configurations defined by
\begin{equation}
 \bm{B}
={B}_{\beta \delta} \ {\bm{A}}^{\beta} \otimes {\bm{A}}^{\delta}\quad \text{and} \quad {\bm{b}}
={b}_{\beta \delta} \ {\bm{a}}^{\beta} \otimes {\bm{a}}^{\delta},   
\end{equation}
where
\begin{equation}
  {B}_{\beta \delta}=\bm{N} \cdot {\bm{A}}_{\beta , \delta},  \quad \text{and} \quad {b}_{\beta \delta}={\bm{n}} \cdot {\bm{a}}_{\beta , \delta},
\end{equation}
and further,
\begin{equation}
 {{B}_\beta}^{\gamma}={B}_{\beta\delta}{A}^{\delta \gamma},\quad \text{and} \quad {{b}_\beta}^{\gamma}={b}_{\beta \delta}{a}^{\delta \gamma}. 
\end{equation}
Therefore,
\begin{equation}
  \bm{K} ={B}_\beta^{ \ \gamma} \ {\bm{A}}_{\gamma}\otimes {\bm{A}}^{\beta}, \quad \text{and} \quad  \bm{\kappa}={{b}}_\beta^{ \ \gamma} \ {\bm{a}}_{\gamma}\otimes {\bm{a}}^{\beta}.
\end{equation}
For a point in the shell, the components of the covariant and contravariant metric tensors in the reference configuration are 
\begin{equation}
{G}_{\alpha \beta} = {\bm{G}}_{\alpha} \cdot {\bm{G}}_{\beta}\quad \text{and} \quad {G}^{\alpha \beta} = {\bm{G}}^{\alpha} \cdot {\bm{G}}^{\beta},
\end{equation}
with their deformed counterparts as
\begin{equation}
    g_{\alpha \beta} = \bm{g}_{\alpha} \cdot \bm{g}_{\beta} \quad \text{and} \quad g^{\alpha \beta} = \bm{g}^{\alpha} \cdot \bm{g}^{\beta},
    \label{eq:contra_metric}
\end{equation}
where ${\bm{G}}^{\alpha}$ and $\bm{g}^{\alpha}$ denotes the contravariant basis vectors in the two configurations defined by
\begin{equation}
{\bm{G}}^{\alpha} \cdot {\bm{G}}
_{\beta} = \delta^\alpha_\beta, \quad \text{and} \quad \bm{g}^{\alpha} \cdot \bm{g}_{\beta} = \delta^\alpha_\beta  .
\label{eq:deform_tensor_1}
\end{equation}
It can be shown that
\begin{equation}
    {\bm{G}}^{\alpha}={\bm{M}}^{-\mathrm{T}} {\bm{A}}^{\alpha},
     \label{def_grad_deriv_2}
\end{equation}
and  ${\bm{M}}^{-1}$ can be expanded as
\begin{eqnarray}
    {\bm{M}}^{-1}
    &=&\left[{M}^{-1\gamma}_{0_\beta} + 
    \eta {M}^{-1  \gamma}_{1_\beta} + 
    \eta^2 {M}^{-1  \gamma}_{2_\beta}\right] {\bm{A}}_{\gamma} \otimes {\bm{A}}^{\beta}+ \bm{\mathcal{O}}(\eta^3),
    \label{def_grad_deriv_1}
\end{eqnarray}
with
\begin{equation}
{M}^{-1\gamma}_{0_\beta}= \delta^\gamma_\beta, \quad
{M}^{-1  \gamma}_{1_\beta}= {B}_\beta^{ \ \gamma}, \quad \text{and} \quad
{M}^{-1  \gamma}_{2_\beta}= {B}_\delta^{ \ \gamma}  {B}_\beta^{ \ \delta}. 
\end{equation}
Similarly,
\begin{equation}
    \bm{g}^{\alpha}=\bm{\mu}^{-\mathrm{T}}\bm{a}^{\alpha},
     \label{def_grad_deriv_3}
\end{equation}
and  $\bm{\mu}^{-1}$ can be expanded as
\begin{equation}
    \bm{\mu}^{-1}
    =\left[{\mu}^{-1  \gamma}_{0_\beta}+ 
    \eta {\mu}^{-1  \gamma}_{1_\beta} + 
    \eta^2 {\mu}^{-1  \gamma}_{2_\beta}\right] {\bm{a}}_{\gamma} \otimes {\bm{a}}^{\beta}+ \bm{\mathcal{O}}(\eta^3),
    \label{def_grad_deriv_4}
\end{equation}
with
\begin{equation}
{\mu}^{-1  \gamma}_{0_\beta}= \delta^\gamma_\beta,\quad
{\mu}^{-1  \gamma}_{1_\beta}= \lambda{b}_\beta^{ \ \gamma}, \quad \text{and} \quad
{\mu}^{-1  \gamma}_{2_\beta}= \lambda^2{b}_\delta^{ \ \gamma} \ {b}_\beta^{ \ \delta}. 
\end{equation}
\subsection{The volume and surface elements}
The volume element in the reference configuration can be expressed as
\begin{eqnarray}
dV=\left[{\bm{G}}_1 \times {\bm{G}}_2\right] \cdot \bm{N} \ d\theta^1 d\theta^2 d\eta
=  \ \left[{\bm{A}}_1 \times {\bm{A}}_2\right] \cdot \bm{N}  M d\theta^1 d\theta^2 d\eta
=dS d\eta,
\label{diff_vol}
\end{eqnarray}
where the undeformed elemental area $dS$ is given by 
\begin{equation}
dS=M d{S}_\mathrm{m} ,
\label{eq:undeformed_area_arb}
\end{equation}
with $d{S}_\mathrm{m}$ as the area element on ${S}_\mathrm{m}$ written as
\begin{equation}
    dS_{\mathrm{m}}={A}^{1/2} dP,
    \label{eq:undeformed_area_mid}
\end{equation}
and the area element for the convected coordinates is $ dP=d\theta^1 d\theta^2$.
Also,
\begin{eqnarray}
M=\mathrm{det}{\bm{M}}
=1-2\eta H+\eta^2 K,
\label{eq:shift_ref}
\end{eqnarray}
where $H$ and $K$ are the mean and Gaussian curvatures of the undeformed mid-surface, respectively, and are expressed as
\begin{equation}
    H=\frac{1}{2}\mathrm{tr}{\bm{K}}=\frac{1}{2}\frac{\partial{\bm{N}}}{\partial {\bm{R}}}:{\bm{I}}=\frac{1}{2}{B}_{ \alpha \beta}{A}^{\alpha \beta}=\frac{1}{2}{B}_\alpha^{\alpha},
\end{equation}
and
\begin{equation}
    K=\mathrm{det}{\bm{K}}
    =\mathrm{det}\left[{B}_\alpha^{ \ \beta}\right]=\mathrm{det}\left[{B}_{ \alpha \gamma}{A}^{\gamma \beta}\right]=\frac{B}{A},
\end{equation}
with $B=\mathrm{det}\left[{B}_{ \alpha \beta}\right]$.
Further, an elemental area in the deformed configuration is given by
\begin{equation}
    ds=\mu {\hat{a}}^{1/2} dS_\mathrm{m},
\end{equation}
with 
the surface stretch $\hat{a}=\dfrac{a}{A}$, 
and
\begin{equation}
    \mu=1-2\eta \lambda h+\eta^2 \lambda^2 \kappa,
\end{equation}
where the mean and Gaussian curvatures of the deformed mid-surface are
\begin{equation}
     h=\frac{1}{2}{b}_\alpha^{\alpha},\quad \text{and} \quad \kappa=\frac{b}{a},
\end{equation}
with $b=\mathrm{det}[b_{ \alpha \beta}]$.
Therefore,
\begin{equation}
   d S_\mathrm{t}={M}\Big|_{\eta=T/2}
   dS_\mathrm{m}, \quad \text{and} \quad  d S_\mathrm{b}={M}\Big|_{\eta=-T/2}
   d{S}_{\mathrm{m}}.
   \label{eq:ref_top_bot_to_mid}
\end{equation}
Also,
\begin{equation}
   d s_\mathrm{t}=\mu\Big|_{\eta=T/2}
   {\hat{a}}^{1/2} d S_\mathrm{m}, \quad \text{and} \quad  d s_\mathrm{b}=\mu\Big|_{\eta=-T/2}
   {\hat{a}}^{1/2} dS_\mathrm{m}.
   \label{eq:def_surface_top_bottom}
\end{equation}
If the bounding curve  $C_\mathrm{m}$ of  the mid-surface $S_\mathrm{m}$  is characterised by the arc-length parameter $l$, then the infinitesimal length $dl$ between two points ${\bm{R}}(\theta^1, \theta^2)$ and ${\bm{R}}(\theta^1 + d\theta^1, \theta^2+d\theta^2)$ is given by
\begin{eqnarray}
dl=\norm{{\bm{R}}(\theta^1 + d\theta^1, \theta^2+d\theta^2)-{\bm{R}}(\theta^1, \theta^2)}
=\norm{{\bm{A}}_\gamma d\theta^\gamma}
= \sqrt{{\bm{A}}_\alpha d\theta^\alpha \cdot {\bm{A}}_\beta d\theta^\beta}
= \sqrt{A_{\alpha \beta}d\theta^\alpha d\theta^\beta}.
\label{eq:dl}
\end{eqnarray}
\begin{figure}
  \begin{center}
   \includegraphics[width=0.6\linewidth]{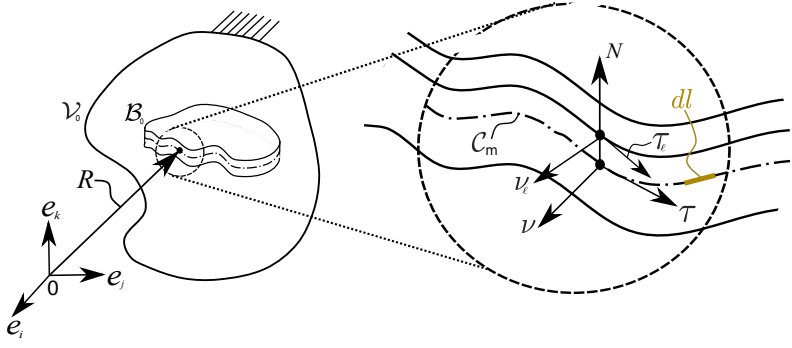}
    \caption{A schematic showing the local triads at any point on the lateral surface and at a point on the bounding curve of the mid-surface in the reference configuration of the shell.}
      \label{fig:metrics_1}
  \end{center}
\end{figure}
The tangent vector at a point $\bm{R}$ on $C_\mathrm{m}$ is defined as 
\begin{equation}
    \bm{\tau}=\frac{d\bm{R}}{dl}=\frac{\partial \bm{R}}{\partial\theta^\beta}\frac{d\theta^\beta}{dl}={\bm{A}}_\beta \frac{d\theta^\beta}{dl},
\end{equation}
and using Equation \eqref{eq:dl}, 
\begin{eqnarray}
  \bm{\tau} \cdot \bm{\tau}={\bm{A}}_\alpha \cdot {\bm{A}}_\beta \frac{d\theta^\alpha}{dl} \frac{d\theta^\beta}{dl}
  ={A}_{\alpha \beta} \frac{d\theta^\alpha}{dl} \frac{d\theta^\beta}{dl}
  =1,  
\end{eqnarray}
implying that $\bm{\tau}$ is a unit tangent vector. Further, define 
\begin{equation}
    \bm{\nu}=\bm{E}\bm{\tau}={E}^{\alpha \beta}{A}_{\beta \gamma}\frac{d\theta^\gamma}{dl} {\bm{A}}_\alpha={E}_{\eta \delta} \frac{d \theta^\delta}{dl} {\bm{A}}^\eta,
    \label{eq:in-plane_normal}
\end{equation}
such that,
\begin{eqnarray}
     \bm{\nu} \cdot \bm{\tau}={E}_{\alpha \beta} \frac{d \theta^\alpha}{dl} \frac{d \theta^\beta}{dl}
     =\frac{1}{2}\left[{E}_{\alpha \beta}+{E}_{\beta \alpha}\right]\frac{d \theta^\alpha}{dl} \frac{d \theta^\beta}{dl}
     =0.
\end{eqnarray}
Again,
\begin{eqnarray}
  \bm{\nu} \cdot  \bm{\nu} = {E}^{\alpha \beta}{A}_{\beta \gamma}\frac{d\theta^\gamma}{dl} {\bm{A}}_\alpha \cdot {E}_{\eta \delta} \frac{d \theta^\delta}{dl} {\bm{A}}^\eta 
  ={E}^{\alpha \beta}{E}_{\eta \delta}\delta^{\eta}_\alpha {A}_{\beta \gamma} \frac{d\theta^\gamma}{dl}\frac{d \theta^\delta}{dl},
\end{eqnarray}
and following the relation,
   $ {E}^{\alpha \beta}{E}_{\eta \delta}\delta^{\eta}_\alpha=\left[\delta^{\alpha}_\eta\delta^{\beta}_\delta-\delta^{\alpha}_\delta\delta^{\beta}_\eta\right]\delta^{\eta}_\alpha
    =\delta^{\beta}_\delta$,
\begin{eqnarray}
  {\bm{\nu}} \cdot \bm{\nu}=\delta^{\beta}_\delta {A}_{\beta \gamma} \frac{d\theta^\gamma}{dl}\frac{d \theta^\delta}{dl}
  ={A}_{\delta \gamma} \frac{d\theta^\gamma}{dl}\frac{d \theta^\delta}{dl}
  = \bm{\tau} \cdot \bm{\tau}
  =1,  
\end{eqnarray}
implying that $\bm{\nu}$ is the in-plane unit normal to $\bm{\tau}$ on $\mathcal{C}_\mathrm{m}$, and
\begin{equation}
    \bm{\nu}=\bm{\tau}\times \bm{N}.
\end{equation}
An elemental area, $d {S}_\ell$, at a point $\bm{X}_\mathrm{B}$ on the lateral surface is given by
\begin{eqnarray}
d S_\ell=\norm{\frac{\partial \bm{X}_\mathrm{B}}{\partial l}\times \frac{\partial \bm{X}_\mathrm{B}}{\partial \eta}}dl d\eta
=\norm{\bm{M}\bm{\tau} \times \bm{N}}dl d\eta
=c\norm{{\bm{\tau}}_\ell\times \bm{\bm{N}}}dl d\eta
=c dl d\eta,
\label{eq:lateral_to_mid}
\end{eqnarray}
with
\begin{equation}
    c=\norm{\bm{M}\bm{\tau}}={\left[1-2\eta \bm{K} \bm{\tau} \cdot \bm{\tau}+\eta^2 \bm{K}\bm{\tau} \cdot \bm{K}\bm{\tau}\right]}^{1/2}, \quad \text{and} \quad \bm{\tau}_\ell=\frac{\bm{M}\bm{\tau}}{c}, 
\end{equation}
where ${\bm{\tau}}_\ell$ is the unit tangent vector at a point on the lateral surface, and   
the in-plane unit normal is
\begin{equation}
    {\bm{\nu}}_\ell={\bm{\tau}}_{\ell}\times\bm{N}.
\end{equation}
The above can be written as
\begin{equation}
     c{\bm{\nu}}_\ell =\left[\bm{I}+\eta\left[\bm{K}-2H\bm{I}\right]\right]\bm{\nu},
     \label{eq:lateral_nor_to_mid}
\end{equation}
by using the relation,
\begin{equation}
     \bm{K}\bm{\tau}\times \bm{N}=\left[2H\bm{I}- 
     \bm{K}\right]\bm{\nu}.
\end{equation}
For clarity, Figure \ref{fig:metrics_1} illustrates the local triads at a point on the bounding curve of the mid-surface and at a shell-point on the lateral surface.
\end{appendices}
\begin{appendices}
\renewcommand{\theequation}{B.\arabic{equation}}
\section{Application of Green's Theorem at the mid-surface of the shell}
For a scalar ${T}^\alpha$, consider the following integral:
\begin{equation}
\int\limits_{ P}\left[{A}^{1/2}{T}^\alpha\right]_{, \alpha}dP,
\end{equation}
which can be rewritten by applying the Green's theorem as
\begin{eqnarray}
  \int\limits_{ P}\left[{A}^{1/2}{T}^\alpha\right]_{, \alpha}dP = \int\limits_{ P}\left[\left[{A}^{1/2}{T}^1\right]_{, 1}+\left[{A}^{1/2}{T}^2\right]_{, 2}\right]dP
  = \int\limits_{\mathcal{C}_\mathrm{p}} {A}^{1/2} {e}_{\alpha \beta}{T}^\alpha d\theta^\beta,
\end{eqnarray}
where $\mathcal{C}_\mathrm{p}$ is the boundary of the parametric domain $P$, and the above boundary integral can be simplified as
\begin{eqnarray}
  \int\limits_{\mathcal{C}_\mathrm{p}} {A}^{1/2} {e}_{\alpha \beta}{T}^\alpha d\theta^\beta
  =\int\limits_{\mathcal{C}_\mathrm{p}} {E}_{\alpha \beta}{T}^\alpha d\theta^\beta
  =\int\limits_{\mathcal{C}_\mathrm{m}}\left[{E}_{\alpha \beta}\frac{d\theta^\beta}{dl}\right]{T}^\alpha dl, 
\end{eqnarray}
which on using Equation \eqref{eq:in-plane_normal} can be further written as
\begin{equation}
 \int\limits_{\mathcal{C}_\mathrm{p}} {A}^{1/2} {e}_{\alpha \beta}{T}^\alpha d\theta^\beta=\int\limits_{\mathcal{C}_\mathrm{m}}{T}^\alpha {\nu}_\alpha dl.   
\end{equation}
This establishes a relation between the integral over the parametric domain and the line integral along the boundary of the curved mid-surface.
\end{appendices}
\begin{appendices}
\renewcommand{\theequation}{B.\arabic{equation}}
\section{Variation of some relevant quantities}
\label{app: variation of relevant quantities}
Here, the first variation of key kinematic variables are listed (essential for the calculations in Sections 5 and 6, respectively),
for example,
\begin{equation}
    \delta\bm{F}=\delta\frac{\partial \bm{\chi}}{\partial \bm{X}}
    =\frac{\partial \delta\bm{\chi}}{\partial \bm{X}},
\end{equation}
and following
$
    \bm{F}\bm{F}^{-1}=\mathbold{\mathbb{1}},
$
one obtains
\begin{equation}
    \delta \bm{F}^{-1}=-\bm{F}^{-1} \delta \bm{F}\bm{F}^{-1}. 
\end{equation}
Also, 
\begin{equation}
    \delta J=J\bm{F}^{-T}:\delta\bm{F}.
    \label{eq:var_jac}
\end{equation}
Again, 
\begin{eqnarray}
    \delta\bm{\chi}_\mathrm{B}&=&\delta \bm{r}+\eta \delta \bm{d}, \nonumber \\
    &=&\delta \bm{r}+\eta \delta\lambda  \bm{n} + \eta  \lambda  \delta\bm{n}, 
    \label{eq:var_chi_1}
\end{eqnarray}
where $\delta \bm{n}$ can be obtained by using the relations, $\bm{n}\cdot\bm{n}=1$ and $\bm{a}_{\alpha}\cdot\bm{n}=0$ as
\begin{equation}
    \delta \bm{n}=-\left[\bm{a}^\alpha \otimes \bm{n}\right]\delta \bm{a}_\alpha=-\bm{a}^\alpha \left[\bm{n} \cdot \delta \bm{a}_\alpha\right],
    \label{eq:var_def_nor}
\end{equation}
with
\begin{equation}
    \delta \bm{a}_\alpha=\delta \frac{\partial \bm{r}}{\partial \theta^\alpha}=\left[\delta \bm{r}\right]_{, \alpha},
\end{equation}
and moreover, from Equation~\eqref{eq:through_thickness} follows
\begin{equation}
    \delta \lambda= -\frac{\lambda}{2} a^{-1}\delta a,
\end{equation}
where from Equation~\eqref{alternate_exp_det},
\begin{equation}
    \delta a=a a^{\alpha \beta}\delta a_{\alpha \beta},
\end{equation}
 which can be rewritten by using
$
    \delta a_{\alpha \beta}=\delta {\bm{a}}_{\alpha} \cdot  {\bm{a}}_{\beta} +  {\bm{a}}_{\alpha} \cdot \delta {\bm{a}}_{\beta}
$
as
\begin{equation}
    \delta a= 2a \bm{a}^{\alpha}\cdot\delta {\bm{a}}_{\alpha}.
\end{equation}
Therefore, the variation in the through-thickness stretch can be rewritten as
\begin{equation}
    \delta \lambda=
    -\lambda \bm{a}^{\alpha}\cdot\delta {\bm{a}}_{\alpha}. 
    \label{eq:var_lambda}
\end{equation}
Apart from the kinematic variables, the variation in the magnetic field vector is given by  
\begin{equation}
    \delta \bm{\bbm{H}}= - \delta \frac{\partial \Phi}{\partial \bm{X}}= -\frac{\partial \delta\Phi}{\partial \bm{X}},
    \label{eq:var_mag_field}
\end{equation}
and further,  in ${\mathcal{B}}_0$,
\begin{equation}
    \delta \Phi=\delta {\Phi}_0 + \eta \delta {\Phi}_1.
\end{equation}
Also, neglecting the higher order terms,
\begin{equation}
   \delta p=\delta p_0+\eta \delta p_1
   .
\end{equation}
\end{appendices}

\bibliography{ref} 
\end{document}